\documentclass{ar2e}
\usepackage{ulem}  
\usepackage{ARAstroBib}
\usepackage{amssymb,amsbsy,psfig}
\newcommand{\kms}{km\,s$^{-1}$}

\newcommand{\farcs}{\mbox{\ensuremath{.\!\!^{\prime\prime}}}}

\newcommand{\apj}{ApJ}%
\newcommand{\apjs}{ApJS}
\newcommand{\apjl}{ApJL}
\newcommand{\aap}{A{\&}A}

\newcommand{\mnras}{MNRAS}
\newcommand{\aj}{AJ}
\newcommand{\araa}{ARAA}
\newcommand{\pasp}{PASP}
\newcommand{\nat}{Nature}
\newcommand{\fdm}{f$_{\rm DM}$}

\begin{document}


\jname{Annu. Rev. Astron. Astrophys.}
\jyear{2010}
\jvol{}
\ARinfo{}

\title{Strong Lensing by Galaxies}

\markboth{Tommaso Treu}{Strong Lensing by Galaxies}

\author{Tommaso Treu
\affiliation{Physics Department, University of California, Santa Barbara CA 93106-9530, tt@physics.ucsb.edu; Sloan Research Fellow; Packard Research Fellow.}}


\begin{abstract}
Strong lensing is a powerful tool to address three major astrophysical
issues: understanding the spatial distribution of mass at kpc and
sub-kpc scale, where baryons and dark matter interact to shape
galaxies as we see them; determining the overall geometry, content,
and kinematics of the universe; studying distant galaxies, black
holes, and active nuclei that are too small or too faint to be
resolved or detected with current instrumentation.  After summarizing
strong gravitational lensing fundamentals, I present a selection of
recent important results.  I conclude by discussing the exciting
prospects of strong gravitational lensing in the next decade.
\end{abstract}

\maketitle

\section{Introduction}

As photons from distant sources travel across the universe to reach
our telescopes and detectors, their trajectories are perturbed by the
inhomogeneous distribution of matter. Most sources appear to us
slightly displaced and distorted in comparison with the way they would
appear in a perfectly homogeneous and isotropic universe. This
phenomenon is called weak gravitational lensing \citep[e.g.][and
references therein]{Ref03}. Under rare circumstances, the deflection
caused by foreground mass overdensities such as galaxies, groups, and
clusters is sufficiently large to create multiple images of the
distant light source. This phenomenon is called strong gravitational
lensing.  Due to space limitations, this article will focus on cases
where gravitational lensing is caused primarily by a galaxy-sized
deflector (or lens).

The first strong gravitational lens was discovered more than thirty
years ago, decades after the phenomenon was predicted theoretically
\citep[see][and references therein]{B+N92}. However, in the past decade 
there has been a dramatic increase in the number of known lenses and
in the quality of the data.  At the time of the review by
\citet{B+N92}, the 11 ``secure'' known galaxy-scale lenses could all
be listed in a page and discussed individually. At the time of this
writing, the number of known galaxy-scale lens systems is
approximately 200, most of which have been discovered as part of large
dedicated surveys with well defined selection functions.  This
breakthrough has completed the transformation of gravitational lensing
from an interesting and elegant curiosity to a powerful tool of
general interest and statistical power.

Three properties make strong gravitational lensing a most useful tool
to measure and understand the universe. Firstly, strong lensing
observables - such as relative positions, flux ratios, and time delays
between multiple images - depend on the gravitational potential of the
foreground galaxy (lens or deflector) and its derivatives.  Secondly,
the lensing observables also depend on the overall geometry of the
universe via angular diameter distances between observer, deflector,
and source. Thirdly, the background source often appears magnified to
the observer, sometimes by more than an order of magnitude. As a
result, gravitational lensing can be used to address three major
astrophysical issues: i) understanding the spatial distribution of
mass at kpc and sub-kpc scale where baryons and DM interact to shape
galaxies as we see them; ii) determining the overall geometry,
content, and kinematics of the universe; iii) studying galaxies, black
holes, and active nuclei that are too small or too faint to be
resolved or detected with current instrumentation.

The topic of strong lensing by galaxies is too vast to be reviewed
entirely in a single Annual Review Article. This article is meant to
provide an overview of a selection of the most compelling and
promising astrophysical applications of strong gravitational lensing
at the time of this writing. The main focus is on recent results
(after $\sim$2005). For each application, I discuss the context,
recent achievements, and future prospects. Of course, lensing is only
one of the tools of the astronomers' trade. When needed, I discuss
scientific results that rely on strong lensing in combination with
other techniques. For every astrophysical problem, I also present a
critical discussion of whether strong gravitational lensing is
competitive with alternative tools.

Excellent reviews and monographs are available to the interested
reader for more details, different points of view, history of strong
lensing, and a complete list of pre-2005 references.  The Saas Fee
Lectures by Schneider, Kochaneck and Wambsganss (2006) provide a
comprehensive and pedagogical treatment of lensing fundamentals,
theory and observations until 2006. Additional information can be
found in the review by \citet{Fal05} and that by
\citet{CSS02}.  The classic monograph by \citet{SEF92} and that by
\citet{PLW01} are essential references for strong gravitational
lensing theory.

This review is organized as follows. First, for convenience of the
reader and to fix the notation and terminology, in \S~\ref{sec:theory}
I give a very brief summary of strong lensing theory. Then, in
\S~\ref{sec:over}, I present an overview of the current observational
landscape. The following four sections cover the main astrophysical
applications of gravitational lensing: ``The mass structure of
galaxies'' (\S~\ref{sec:mass}), ``Substructure in galaxies''
(\S~\ref{sec:sub}), ``Cosmography'' (\S~\ref{sec:cosmo}), and ``Lenses
as cosmic telescopes'' (\S~\ref{sec:telescopes}). After the four main
sections, the readers left with an appetite for more results from
strong gravitational lensing will be happy to learn about the many
promising ongoing and future searches for more gravitational lenses
described in \S~\ref{sec:searches}. Some considerations on the future
of strong gravitational lensing - when the number of known systems
should be well into the thousands -- are given in~\S~\ref{sec:future}.

\section{Brief Theoretical Introduction}
\label{sec:theory}

\subsection{A gravitational optics primer}

Under standard conditions of a thin lens (i.e. the size of the
deflector is much smaller then the distances between the deflector and
the observer and the deflector and the source), responsible for a weak
gravitational field (i.e. deflection angles much smaller than unity),
in an otherwise homogeneous universe, strong lensing by galaxies can
be described as a transformation from the two-dimensional observed
coordinates associated with a particular light ray (${\boldsymbol
\theta}$ in the image plane) to the two-dimensional coordinates that
the light ray would be observed at in the absence of the deflector
(${\boldsymbol \beta}$ in the source plane).

A simple and intuitive understanding of the basic principles of strong
lensing by galaxies can be gained by considering a generalized version
of Fermat's principle \citep[][and references therein]{B+N92}. For a
given source position ${\bf \beta}$, the excess time-delay surface as
a function of position in the image plane is given by:

\begin{equation}
 t=\frac{D_d D_s (1+z_l)}{cD_{ds}} \left(\frac{1}{2}|{\boldsymbol
 \theta}-{\boldsymbol \beta}|^2- \psi({\boldsymbol \theta}) \right),
\label{eq:fermat}
\end{equation}

\noindent
where $D_d$, $D_s$, $D_{ds}$ are respectively the angular diameter
distances between the observer and the deflector, the observer and the
source, and the deflector and the source, and $\psi$ is the
two-dimensional lensing potential, satisfying the two-dimensional
Poisson Equation:
\begin{equation}
\nabla^2 \psi = 2 \kappa,
\label{eq:poisson}
\end{equation}
\noindent
where $\kappa$ is the surface (projected) mass density of the
deflector in units of the critical density $\Sigma_c=c^2 D_s / (4 \pi
G D_d D_{ds})$
\citep[for convenience of the reader I adopt the same notation
as][]{SKW06}.

\begin{figure}
\centerline{\psfig{file=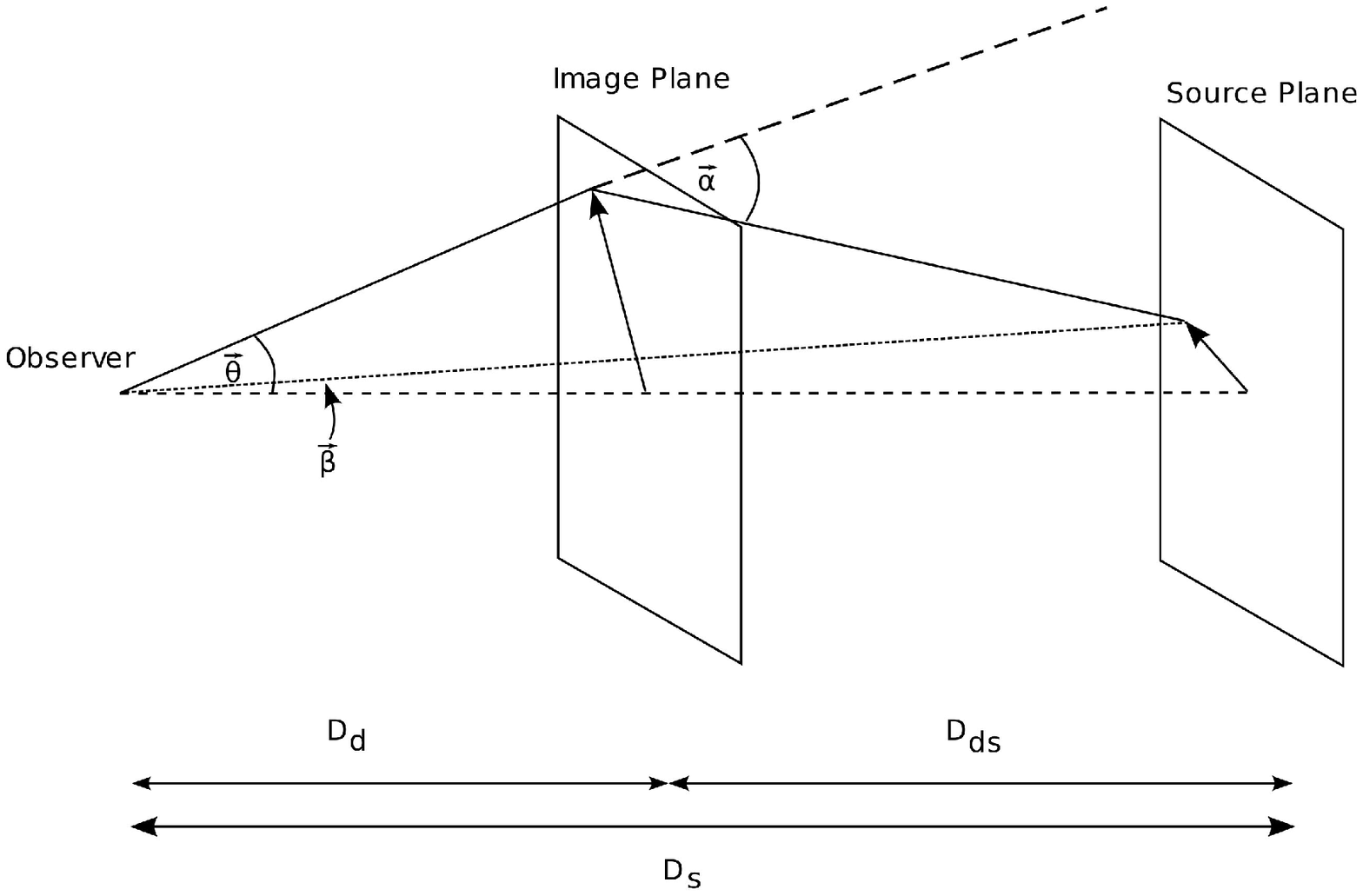,width=\textwidth}}
\caption{Sketch of the gravitational lensing geometry, courtesy of B.Brewer.}
\label{fig:lensgeom}
\end{figure}

According to Fermat's principle, images will form at the extrema of
the time-delay surface, i.e. at the solutions of the so-called lens
equation:
\begin{equation}
\boldsymbol \beta=\boldsymbol \theta-\nabla \psi = \boldsymbol \theta - \boldsymbol \alpha
\label{eq:lensequation}
\end{equation}
which is the desired transformation from the image plane to the source
plane. The scaled deflection angle $\boldsymbol \alpha$ is related to the
deflection angle experienced by a light ray $\hat{\boldsymbol \alpha}$ by
$\boldsymbol \alpha=\frac{D_{ds} \hat{\boldsymbol \alpha}}{D_s}$. The lensing
geometry is illustrated in Figure~\ref{fig:lensgeom}.  Note that the
transformation is achromatic and preserves surface brightness.

\begin{figure}
\centerline{\psfig{file=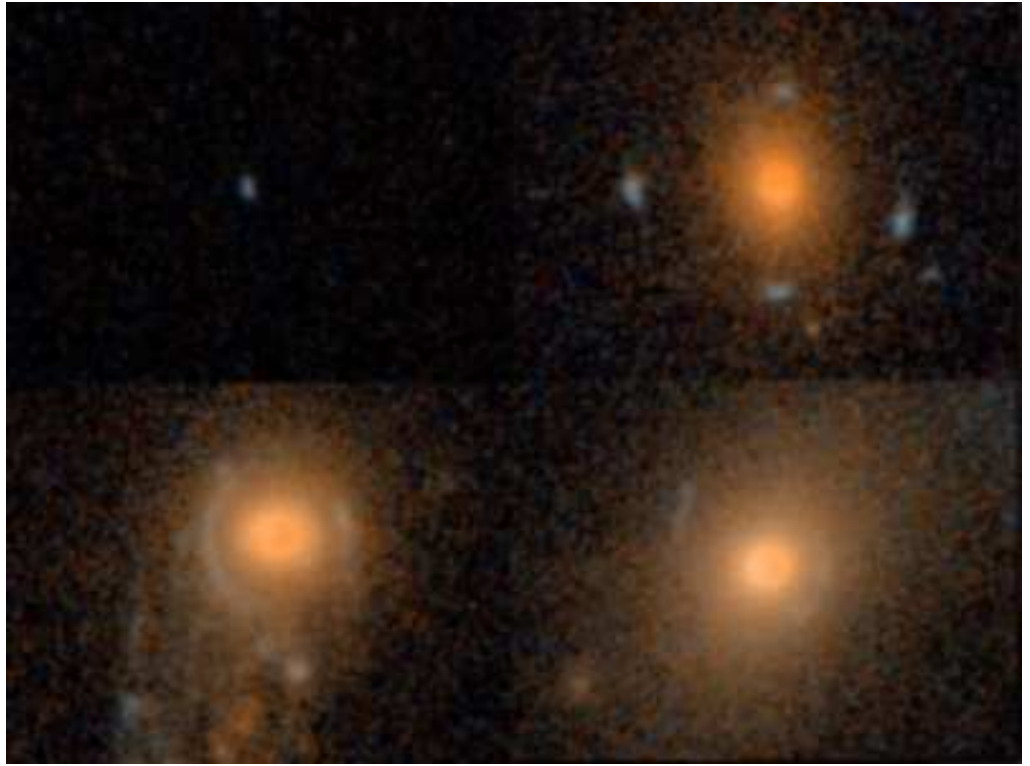,width=\textwidth}}
\caption{Examples of the most common configurations of
galaxy-scale gravitational lens systems.  A background source (top
left) can produce four visible images (a ``quad''; top right), an
(incomplete) Einstein ring (bottom left), or two visible images (a
``double''; bottom right), depending on the ellipticity of the
projected mass distribution of the deflector and on the relative
alignment between source and deflector \citep[data from][Image
courtesy of P.~Marshall]{Mou++07}.}
\label{fig:example}
\end{figure}

\begin{figure}
\centerline{\psfig{file=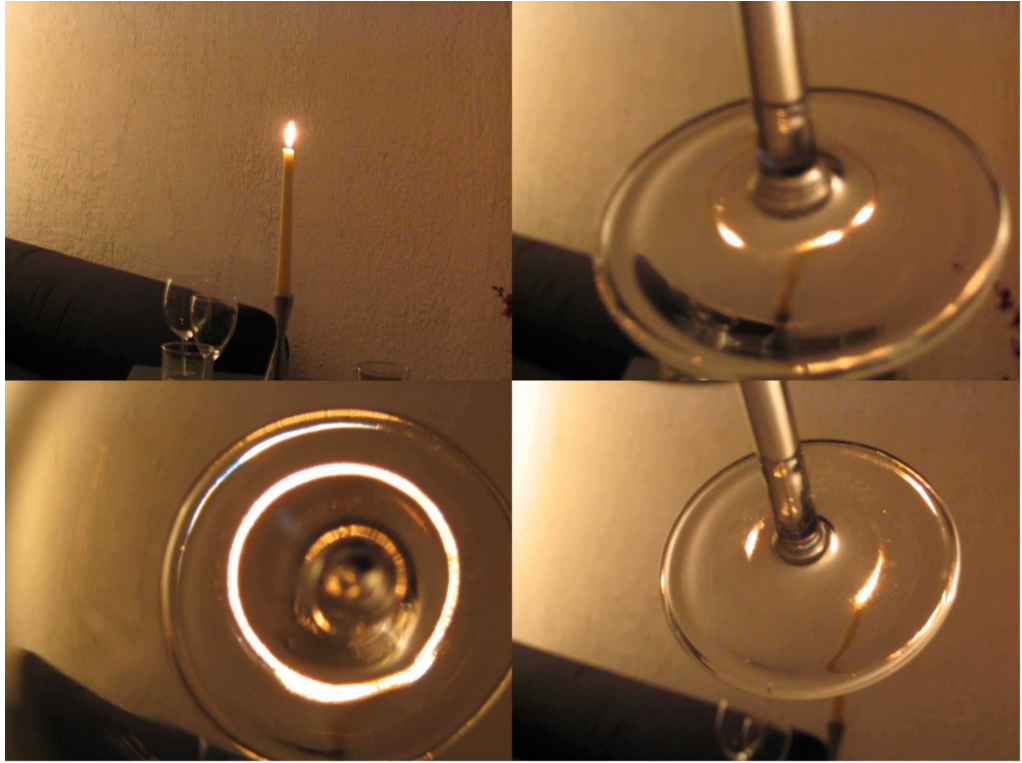,width=\textwidth}}
\caption{Optical analogy to illustrate the gravitational lensing
phenomenon. The optical properties of the stem of a wineglass are
similar to those of a typical galaxy scale lens. Viewed through a
wineglass, a background compact source such as distant candle (top
left), can reproduce the quad (top right), Einstein ring (bottom
left), and double (bottom rights) configurations observed in
gravitational lensing and shown in Figure~\ref{fig:example}. Image
courtesy of P.~Marshall.}
\label{fig:analogy}
\end{figure}

Strong lensing occurs when Equation~\ref{eq:lensequation} has multiple
solutions corresponding to multiple images. Examples of the most
common configurations of strong gravitational lensing by galaxies are
shown in Figure~\ref{fig:example} and explained with an optical
analogy in Figure~\ref{fig:analogy}. For a given deflector the solid
angle in the source plane that produces multiple images is called the
strong-lensing cross section. For a given population of deflectors,
the optical depth is the fraction of the sky where distant sources
appear to be multiply-imaged.

The Jacobian of the transformation from the image to the source plane,
gives the inverse magnification tensor, which can be written as
\begin{equation}
\frac{\partial \boldsymbol \beta}{\partial \boldsymbol \theta}=\delta_{ij}-\frac{\partial^2 \psi}{\partial \theta_i \partial \theta_j}=\left(\begin{array}{cc} 1-\kappa-\gamma_1 & -\gamma_2 \\ 
-\gamma_2 & 1-\kappa+\gamma_1 \end{array} \right)
\label{eq:magnification}
\end{equation}
and describes the local isotropic magnification of a source
(determined by the convergence $\kappa$, defined above) and its
distortion (shear components $\gamma_1$, and $\gamma_2$). 

In the limit of a point source, the local magnification $\mu$ is given
by the determinant of the magnification tensor:

\begin{equation}
\mu=\frac{1}{(1-\kappa)^2-\gamma_1^2-\gamma_2^2}.
\label{eq:scalarmag}
\end{equation}
\noindent
For extended sources, the observed magnification depends on the surface
brightness distribution of the source as well as on the magnification
matrix.

When the determinant of the inverse magnification matrix vanishes, the
magnification becomes formally infinite. The loci of formally infinite
magnification in the image plane are called critical lines. The
corresponding loci in the source plane are called caustics. Compact
sources located close to a caustic can be magnified by very large
factors up to almost two orders of magnitude \citep{Sta++08}, although
the total observed flux is always finite for astrophysical sources of
finite angular size.

It is convenient to define the Einstein radius. For a circular
deflector it is the radius of the region inside which the average
surface-mass density equals the critical density. A point source
perfectly aligned with the center of a circular mass distribution will
be lensed into a circle of radius equal to the Einstein radius, the
so-called Einstein ring (see Figure~\ref{fig:example}). The size of
the Einstein radius depends on the enclosed mass as well as on the
redshifts of deflector and source. The definition of Einstein radius
needs to be modified for non-circular deflectors \citep{KSB94}. Once
appropriately defined, the Einstein radius is a most useful quantity
to express the lensing strength of an object, and it is usually very
robustly determined via strong lens models \citep[e.g.][]{SKW06}. As a
consequence, the mass enclosed in the cylinder of radius equal to the
Einstein radius can be measured to within 1-2\%, including all random
and systematic uncertainties.

A final essential concept is that of mass-sheet degeneracy
\citep{FGS85}. Given dimensionless observables in the image plane,
such as relative position, shape, and flux ratios of multiple images,
the solution of the lens equation is not unique. For every mass
distribution $\kappa(\boldsymbol \theta)$ and every surface brightness
distribution in the source plane I($\boldsymbol \beta$), there is a
family of solutions given by the transformations:
\begin{equation}
\kappa_{\lambda}=(1-\lambda)+\lambda \kappa ; \quad \boldsymbol \beta_\lambda = \boldsymbol \beta/\lambda
\label{eq:masssheet}.
\end{equation}
The transformation changes the predicted time-delay between multiple
images and the magnification as follows:
\begin{equation}
\Delta t_{\lambda}=\Delta t /\lambda ; \quad
\mu_\lambda=\mu/\lambda^2,
\label{eq:msscaling}
\end{equation}
resulting in a degeneracy in inferred quantities such as intrinsic
luminosity and size of the background source. Additional information
is needed to break this degeneracy, such as the intrinsic luminosity
or size of the lensed source \citep[as in the case of lensed
supernovae Ia][]{K+B98}, the actual mass of the deflector (as measured
for example with stellar kinematics), or the measured time-delays
between multiple images within the context of fixed cosmology.
Alternatively, the mass-sheet degeneracy can be broken in the context
of a model, for example by assuming that the surface mass density
of the deflector goes to zero at large radii (thus $\lambda=1$). In
practice, this is not always possible because mass structure along the
line of sight - associated or not with the main deflector -- can act
effectively as a ``sheet'' of mass with external convergence
$\kappa_{\rm ext}$. Breaking the mass-sheet degeneracy is essential
for a number of strong lensing applications, as we will see in
Section~\ref{sec:cosmo}.

\subsection{Modeling galaxies: macro, milli, and micro lensing}

\begin{figure}
\centerline{\psfig{file=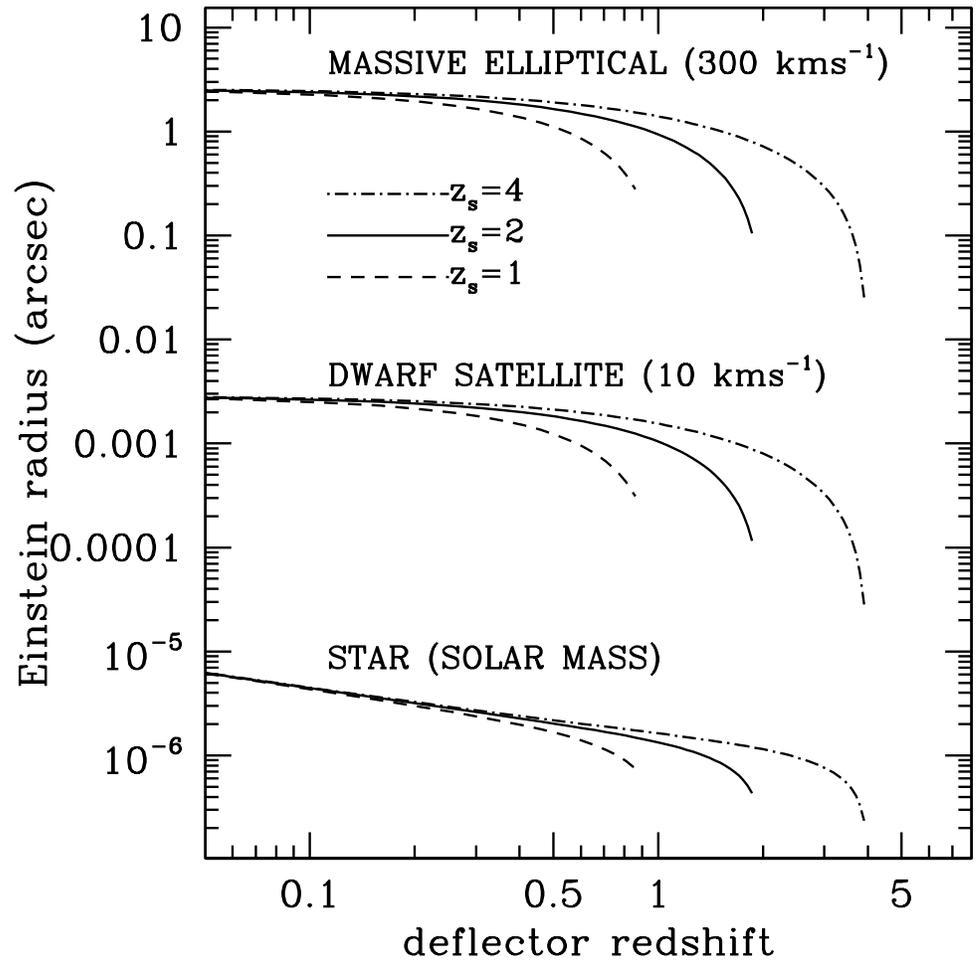,width=\textwidth}}
\caption{Einstein radius of a massive elliptical galaxy (top), a 
dwarf satellite (middle) and a star (bottom) as a function of
deflector redshift for three choices of source redshifts
($z_s=1,2,8$). SIS models with velocity dispersion $\sigma$=300 and 10
kms$^{-1}$ are assumed for the elliptical and dwarf galaxies
respectively. A point mass of one solar mass is adopted for the star.}
\label{fig:lens}
\end{figure}

It is useful to define three regimes to describe the lensing
properties of the components of galaxies, corresponding to the typical
scale of associated Einstein radii, as summarized in
Figure~\ref{fig:lens}.

\subsubsection{Macrolensing}

On the coarsest scale, corresponding to Einstein radii of the order of
arcseconds, the overall mass distribution of the lensing galaxy is
responsible for the main features of the multiple images, such as
image separation and multiplicity. In terms of physical components of
an isolated galaxy, macrolensing can be thought of as the combined
lensing properties of the DM halo, the bulge, and the disk. A simple
model that reproduces image positions, multiplicity, and fluxes is
sometimes referred to as the macro model, and is generally sufficient
to infer quantities such as projected mass inside the Einstein radius
and overall ellipticity and orientation of the mass distribution.  The
simplest model that is found to provide a qualitatively good
description of the macroscopic features of strong lensing by galaxies
is the SIE \citep{KSB94}, an elliptical generalization of the SIS. The
three-dimensional mass density profile of the SIS is given by

\begin{equation}
\rho=\frac{\sigma_{\rm SIS}^2}{2\pi G r^2},
\label{eq:SIS}
\end{equation}
and the Einstein radius is given by 

\begin{equation}
\theta_E=4\pi\left(\frac{\sigma_{\rm
SIS}}{c}\right)^2\frac{D_{ds}}{D_s}.
\label{eq:thetaSIS}
\end{equation}
\noindent
Note that for early-type lens galaxies $\sigma_{\rm SIS}$ is found to
be approximately equal to the stellar velocity dispersion
\citep[e.g.,][]{Bol++08a}.

\begin{figure}
\centerline{\psfig{file=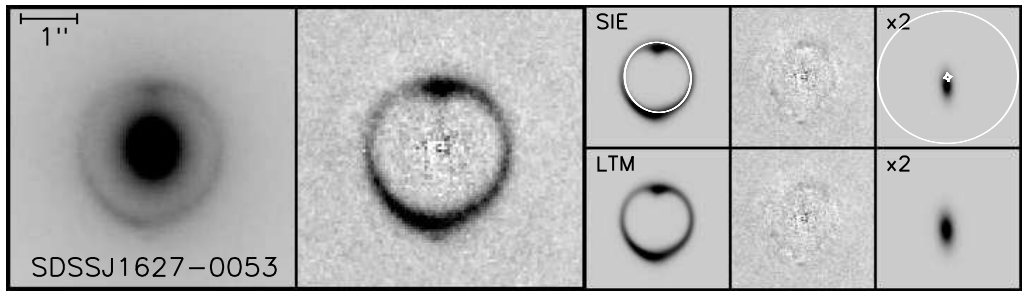,width=\textwidth}}
\caption{Example of a gravitational lens model, from \citet[][reproduced by permission of the AAS]{Bol++08a}.
The two left panels show the data before and after subtraction of the
light from the lens galaxy. The smaller panels on the right show the
predicted image intensity of the best fit lens model, residuals, and
source plane reconstruction, for an SIE mass model (top panels) and a
mass traces light model (bottom panels). In the panel representing the
image plane (labelled SIE) the white line shows the critical line. In
the panel representing the source plane (magnified by a factor of 2)
the white lines show the inner and outer caustics. Note that the peak
of the surface brightness distribution is located outside the inner
caustic and is therefore imaged twice, while the outer regions of the
lensed sources go through the central region and therefore form an
Einstein ring in the image plane.}
\label{fig:boltonmodel}
\end{figure}

An example lens model is shown in Figure~\ref{fig:boltonmodel}. This
system consists of a foreground elliptical galaxy lensing a background
galaxy, well described by an elliptical Gaussian surface brightness
distribution in the source plane. A SIE mass model is found to be
sufficient to reproduce accurately the observed surface brightness
distribution in the image plane. For an SIE mass model, two curves
(outlined in white in the figure) separate regions of different
multiplicity in the source plane. Sources outside the outer curve
(known as cut) are singly imaged, sources in between the cut and the
inner caustic curve produce two visible images (plus a third
infinitely demagnified central image), and sources inside the inner
caustic produce four visible images (plus a fifth infinitely
demagnified central image). In this case, the detectable part of the
extended source crosses the inner caustic so that it appears partly
doubly-imaged and partly quadruply imaged. Due to the good alignment
of the deflector and source, the image forms an almost perfect
Einstein ring. An alternative light traces mass (LTM) model (i.e. the
surface mass density of the deflector is obtained by multiplying its
surface brightness by a mass-to-light ratio, allowed to be a
free-parameter) is also shown. In this case, the LTM model is almost
indistinguishable from an SIE model, because strong lensing is
sensitive only to the mass enclosed by the Einstein radius, to first
order. In general, LTM models can be excluded when considering
extended sources because they fail to reproduce the detailed geometry,
the radial behavior in particular. LTM models can also be excluded on
the basis of a number of other considerations, as reviewed in
\S~\ref{sec:mass}.

We just discussed an example of a simply-parametrized gravitational
lens macro model, where both the source surface brightness and the
mass distribution of the deflector are described by astrophysically
motivated models with a small number of parameters. This kind of model
is generally capable of reproducing all the macroscopic features while
delivering robust estimates of the most important quantities for the
deflector (e.g. total mass ellipticity and orientation) and the source
(e.g. intrinsic size and luminosity). For these reasons, simply
parametrized models are often all one needs in interpreting lensing
data.

However, some applications require more sophisticated lens models,
capable of extracting more detailed information.  In recent years, the
increase in number of known lenses has been paralleled by ever more
sophisticated lens modeling tools.  A full description of advanced
lens models is beyond the scope of this review. However, I list a few
examples to point the interested reader towards the technical
literature.  A number of groups have developed ``grid-based'' models
(also known -- incorrectly -- as non-parametric models; pixel values
are parameters like any other), where the potential (or surface mass
density) of the deflector and/or the surface brightness of the source
are described by a set of pixels on regular or irregular grids, using
regularization schemes to suppress spurious features due to noise
\citep[e.g.][]{W+D03,T+K04,Koo05,D+W05,Suy++06,B+L06b,V+K09a}. An
alternative hybrid approach consists of using large numbers of
simply-parametrized models to strike a balance between flexibility and
prior information on the shape and surface brightness of galaxies
\citep{Mar06}.  Bayesian statistics has become the standard
statistical framework for advanced models, allowing for rigorous
analysis of the uncertainties in highly dimensional spaces as well as
quantitative model selection. Heuristic pixellated approaches have
also been adopted with some success \citep{S+W04}, and recently been
cast in a Bayesian framework to improve the understanding of
the uncertainties \citep[][]{Col08}.

\subsubsection{Millilensing}

On an intermediate angular scale are the lensing effects introduced by
substructure, both luminous and dark. Typically, a lens galaxy will
have some satellites, like the dwarf satellites of the Milky Way
\citep[][and references therein]{Kra10}. The mass associated with the
satellites introduces perturbations in an otherwise smooth
potential. These perturbations can be detected relative to a smooth
model using accurate measurements of flux ratios, relative position,
and time delays between multiple images.  This regime is sometimes
referred to as millilensing -- due to the characteristic
milli-arcsecond Einstein radii expected for dwarf satellites of
massive galaxies. However, the phenomenon could span several orders of
magnitude, depending on the mass function of satellites and their
spatial distribution \citep[e.g.][]{Kra10}.

\subsubsection{Microlensing}

\begin{figure}
\centerline{\psfig{file=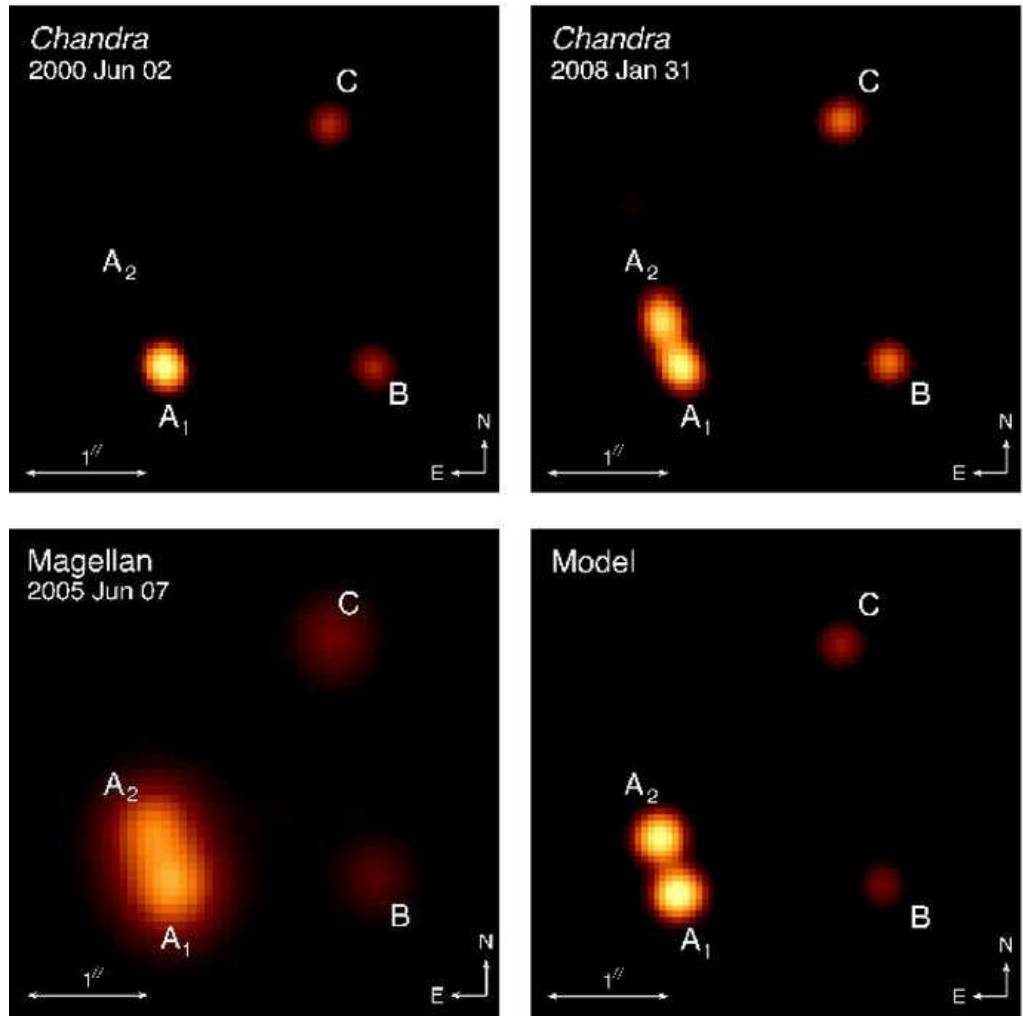,width=\textwidth}}
\caption{Microlensing observed in the quadruply-imaged quasar
PG1115+080 ($z_s=1.72$). The lens galaxy ($z_d=0.31$) has been removed
for clarity. Each panel is 4 arcseconds on a side. The bottom right
panel (labelled Model) shows the expected image predicted from an SIS
model of the deflector and an external shear term to account for the
effects of a nearby group. The flux of image A$_2$ increased by over a
factor of four between June 2000 and January 2008
\citep[Figure from][reproduced by permission of the AAS]{Poo++09}.}
\label{fig:micro}
\end{figure}

Finally, on the smallest angular scale, galaxies are made of
stars. The Einstein radius of a solar mass star at a cosmological
distance is of the order of micro arcseconds, hence the name
cosmological microlensing. The average projected separation of stars
in distant galaxies is small compared to the typical Einstein radii,
and thus every background source effectively experiences cosmological
microlensing.  As in the case of galactic microlensing, the resolution
of current instruments is insufficient to detect cosmological
microlensing via astrometric effects. However, if the angular size of
the background source is smaller or comparable to the typical stellar
Einstein radius, cosmological microlensing can be detected by its
effect on the observed flux. In contrast, if the source is much larger
than the typical stellar Einstein radius, the total magnification will
be effectively averaged over a large portion of the magnification
pattern and therefore be similar to that expected for a smooth
model. The relative motion of stars with respect to the background
source and center of mass of the deflector are sufficiently fast to
modify the magnification pattern over timescales of just a few years,
as illustrated in Figure~\ref{fig:micro}.

As we will see in the rest of this article, all three regimes can be
used to infer unique information on the distribution of mass in
(deflector) galaxies, and on the surface brightness distribution of
distant (lensed) galaxies and active galactic nuclei with sensitivity
and resolution beyond those attainable without the aid of
gravitational lensing.

\section{Observational Overview}
\label{sec:over}

\subsection{Present-day samples and challenges}

\begin{figure}
\centerline{\psfig{file=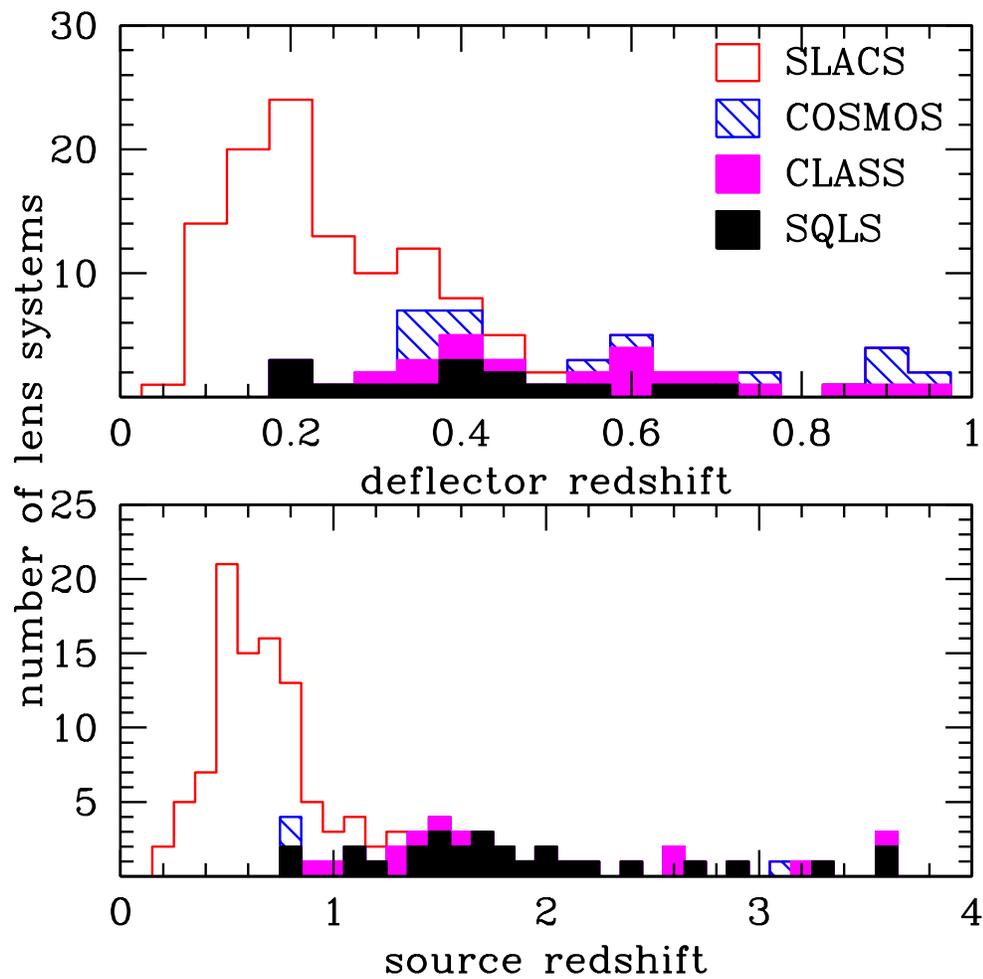,width=\textwidth}}
\caption{Distribution of deflector (top) and source (bottom) spectroscopic
redshifts for the galaxy-scale gravitational lens systems discovered
by the CLASS, COSMOS, SLACS, and SQLS surveys (see text for
details). }
\label{fig:histoz}
\end{figure}

Approximately 200 examples of strong gravitational lensing by galaxies
have been discovered to date. A number of different strategies have
been followed. The two most common strategies start from a list of
potential sources or potential deflectors and use additional
information to identify the (small) subset of strong gravitational
lensing events. Other promising approaches include searching for
gravitational lensing morphologies in high resolution data
\citep[][and references therein]{Mar++09} and exploiting variability 
in time domain data \citep{Koc++06b}. The current state of the art is
illustrated in Figure~\ref{fig:histoz}, which shows the redshift
distribution of the lenses discovered by the four largest surveys to
date. The first two are source-based surveys, the third is a
deflector-based survey, and the fourth one is a lensing morphology
survey.

The Cosmic Lens All-Sky Survey (CLASS) is based on radio imaging. They
discovered 22 multiply-imaged active nuclei, including a subset of 13
systems which are known as the statistically well-defined sample
\citep{Bro++03}. Source and deflector redshifts are available for 11
and 17 systems, respectively (C.D.Fassnacht, 2009 priv. comm). The
SDSS Quasar Lens Search (SQLS), identified 28 galaxy-scale
multiply-imaged quasars using SDSS multicolor imaging data to sift
through the spectroscopic quasar sample
\citep{Ogu++06,Ogu++08}. All source redshifts are available, while
deflector redshifts are available for 15 systems.  The SLACS Survey
\citep{Bol++06} is an optical survey based on spectroscopic
preselection from SDSS data and imaging confirmation with HST. SLACS
discovered 85 galaxies acting as strong lenses \citep[plus an
additional 13 probable lenses;][]{Aug++09a}. Source and deflector
redshifts are available for all systems. Finally, twenty secure
galaxy-scale lens systems were discovered by visual inspection
\citep{Fau++08,Jac08} of the HST images taken as part of the COSMOS
Survey. Source and deflector redshifts are available for 3 and 13
systems, respectively \citep{Lag++09}.

The compilation is not complete, due to the difficulty of keeping
track of the ever growing number of lenses discovered serendipitously
or by ongoing concerted efforts \citep{Cab++07,Mar++09} that still
lack confirmation and spectroscopic redshifts (a useful resource to
find data for lenses from a variety of sources is the online database
of strong gravitational lenses CASTLES at URL
http://www.cfa.harvard.edu/castles).  However, the compilation gives a
good idea of the observational landscape and of the two main
limitations of current samples. Firstly, most new lenses have been
found at $z\lesssim0.4$, which is a very favorable regime for detailed
follow-up, but limits the look-back time baseline for evolutionary
studies and the spatial scales probed by lensing. Secondly, many
gravitational lens systems still lack source or deflector redshifts.

It is customary to classify strong lenses as galaxy-galaxy lenses
(e.g. Figs~\ref{fig:example} and \ref{fig:boltonmodel}), and
galaxy-QSO lenses (e.g. Figure~\ref{fig:micro}), depending on whether
an active galactic nucleus is present in the background source.
Galaxy-QSO lenses are more rare on the sky than galaxy-galaxy lenses
\citep{MBS05}. However, they can be found efficiently by exploiting
their radio emission and the variability of the point
source. Furthermore, the compact point source enables studies of the
granularity of the lens galaxy (from microlensing), and of cosmography
and lens galaxy structure (from direct measurements of time delays
between images). Galaxy-galaxy lenses are typically more suited for
the study of the lens galaxy itself, because its emission is not
overwhelmed by the multiple images of the background
source. Furthermore, the extended surface brightness of the source
provides detailed information on the gravitational potential of the
deflector.

It is observationally challenging to extract the wealth of information
available from strong lensing systems. First and foremost,
subarcsecond angular resolution is key to identifying and
characterizing strong lensing systems. Radio or optical/near infrared
observations from space (and recently from the ground with adaptive
optics) have been essential for the progress of the field. Secondly,
both source and deflector redshifts are needed to transform angular
quantities into masses and lengths. Especially for the source
redshift, long exposure on the largest telescopes are typically
required \citep[e.g.][]{Ofe++06}.  Success is not assured, and in many
cases one must rely on photometric redshifts, which are also
challenging because light from the foreground deflector complicates
photometry of the background source. Third, microlensing and
variability depend critically on source size. This makes X-ray
\citep[e.g.,][]{Poo++09} and 
mid-infrared observations \citep[e.g.,][]{Ago++09} -- probing sources
that are much smaller and much larger than the scale of microlensing,
respectively -- particularly useful, even with limited spatial
resolution. Fourth, time delays and microlensing studies require
intensive monitoring campaigns, with all the associated logistical
challenges. Lastly, depending on the application, ancillary data such
as velocity dispersion or information on the local large scale
structures are typically needed to break degeneracies and control
systematic errors.

\subsection{Selection}

Strong lensing is a very rare phenomenon. With present technology only
$\sim$1/1000 galaxies can be detected as strong lenses
\citep{MBS05}. Similarly, the optical depth is of order
10$^{-3}$-10$^{4}$, i.e.  $\lesssim$1/1000 high-redshift sources in
the sky have detectable multiple images \citep[e.g.][]{Bro++03}. Both
numbers depend strongly on the depth of the observations. Thus, in
order to generalize the results obtained from this technique to the
overall population of deflectors and sources, and for applications of
strong lensing to cosmography, it is essential to understand the
selection function very well.

To first order, strong lensing galaxies can be described as selected
by velocity dispersion. Most galaxy-scale strong gravitational lenses
discovered to date are massive elliptical galaxies with velocity
dispersions in the range 200-300 \kms. This well-understood selection
function arises from the rapid increase in the strong lensing cross
section with mass ($\propto \sigma^4$ for an SIS), and from the rapid
decline of the velocity dispersion function of galaxies above 300
\kms\, \citep[see][for a 
comprehensive discussion]{SKW06}. As an example, the average stellar
velocity dispersion of the SLACS sample is 248 \kms\, with a
r.m.s. scatter of 46 \kms. The velocity dispersion selection is also
responsible for the adverse selection against late-type galaxies.
Approximately 80\% of the SLACS deflectors are pure ellipticals, 10\%
are lenticulars and 10\% are spirals, mostly bulge dominated
\citep{Aug++09a}. Identifying and studying galaxies with
$\sigma\lesssim200$ \kms\, acting as strong gravitational lenses is
possible with sufficiently large surveys and represents an exciting
frontier for the next decade. However, this is an observationally
challenging problem because the image separation drops quickly below
$0\farcs3-0\farcs4$, the current practical limit for detection with
HST and the Very Large Array (VLA). Furthermore, once the resolution
drops significantly below the typical arcsecond size of distant
galaxies, disentangling light from the deflector and background source
becomes increasingly difficult, particularly at optical/infrared
wavelengths.

The lensed sources are to first order flux and surface brightness
selected. This translates into a complex selection function in terms
of the intrinsic properties of the source population because of the
magnification effects of lensing. It is easier to understand the
effect for point source surveys, such as CLASS and SQLS. Due to
lensing magnification, sources that are fainter than the survey flux
limit will enter the sample. However, magnification also reduces the
solid angle actually surveyed. Therefore, the number of strong lensing
events depends critically on the dependency of the surface density of
sources on the observed flux. This effect is known as magnification
bias. For extended sources, observed magnification will also depend on
surface brightness and size of the source, generally being larger for
more compact sources. The redshift distribution of the lensed sources
will in general be different than that for a non-lensed population
selected to the same apparent magnitude limit.

Other more subtle selection effects are also at work. Factors that may
affect the strong lensing cross-section of a galaxy include elongation
along the line of sight, flattening of the projected mass
distribution, concentration of the mass distribution (e.g., the slope
of the mass density profile at fixed virial mass), overdensity of the
local environment, and abundance of small scale structure in the plane
of the deflector or along the line of sight.  Factors that may affect
the probability of a source being identified as multiply-imaged
include extinction from the foreground lens galaxy, configuration of
the multiple images (in particular image separation and flux ratios),
time variability, and presence of emission lines and hence properties
of the stellar populations or existence of an active nucleus.

Three complementary strategies have been followed to quantify
selection effects. One strategy consists of starting from a realistic
cosmological model and simulating the selection process from first
principles \citep[e.g.][and references therein]{MvK09}. This is the
most direct way to compare observations with theoretical models of
galaxy formation. The challenge of this approach is that lensing
selection depends on the details of the mass and surface brightness
distributions on scales much smaller than a galaxy. Unfortunately,
realistic simulations of the universe on this scale are beyond our
current capabilities. Therefore, one needs to rely on DM-only
simulations and approximate the effects of baryons, with all
associated uncertainties. A second strategy consists of comparing
samples of lens galaxies with control samples of non-lens
galaxies. This approach was used with the SLACS sample to show that --
once velocity dispersion and redshifts are matched -- lens galaxies
are indistinguishable within the uncertainties from twin galaxies
selected from SDSS in terms of their size, surface brightness,
luminosity, location on the fundamental plane, stellar mass, and local
environment
\citep{Tre++06,Bol++08a,Tre++09,Aug++09a}. This finding implies that
the results from the SLACS Survey can be applied to the overall
population of velocity dispersion selected early-type galaxies. The
strength of this method is its ability to take into account real
selection functions with all the inherent complexity. This guarantees
that one compares apples with apples, but does not solve the problem
of comparing with theoretical models. A ``hybrid'' approach consists
of constructing simple models starting from empirically-based
information on the deflector and source populations, and combining it
with lensing theory to compute the relevant selection function. This
approach is extremely useful for developing an intuition for the
process and compute approximate correction factors. For example,
\citet{Ogu07} was able to explain the observed ratio of quadruply
imaged to doubly imaged quasars in the CLASS sample in terms of
magnification bias. The challenge for this approach is including a
sufficiently accurate description of the physics and details of the
observations to infer quantitatively correct answers.

\section{The Mass Structure of Galaxies}
\label{sec:mass}

The standard cosmological model, based on CDM and dark energy
reproduces very well the observed structure of the universe on
supergalactic scales \citep[e.g.,][and references therein]{Kom++09}.
At galaxy scales, DM and baryons interact to produce the observed
variety of galaxy properties. The situation is not so clear at small
sub-galactic scales, where potential conflicts between theory and
observations have been suggested \citep[e.g.][]{E+S07}.  Understanding
the interplay between DM and baryons is crucial to make progress in
developing and testing theories of galaxy formation at these
scales. Gravitational lensing, by providing direct and precise
measurements of mass at galactic and subgalactic scales, is a
fundamental tool for answering a number of questions with profound
implications on the existence and nature of DM. Do galaxies reside in
DM halos? How do the properties of galaxies depend on those of their
DM halos?  Are DM density profiles universal as predicted by
simulations? These are the topics of this section.

\subsection{Luminous and dark matter in early-type galaxies}

\subsubsection{Do early-type galaxies live in dark matter halos?}

It is commonly believed that all galaxies live in DM
halos. However, in the case of early-type galaxies, observational
evidence is hard to obtain. The difficulty arises mostly from the
paucity of mass tracers at radii much larger than the effective radius
R$_{\rm e}$ -- where DM dominates -- and from the
degeneracies inherent in interpreting projected data in terms of a
three-dimensional mass distribution for pressure supported systems.
Chief among these degeneracies is that between the total mass density
profile and the anisotropy of the pressure tensor
\citep[``mass-anisotropy'' degeneracy, e.g.,][]{T+K02a}.

Much progress in detecting DM halos has been achieved by studying the
kinematics of stars, globular clusters, and cold and hot gas in nearby
systems \citep[e.g.,][]{B+S93,Hum++06}. This type of study shows that
DM halos are generally required to explain the dynamics of massive
early-type galaxies. Weak-lensing has been used to demonstrate the
existence and to characterize the outer regions of DM halos for
statistical samples of early-type galaxies out to intermediate
redshifts \citep[$z\sim0.5$, e.g.][]{Lag++09,Hoe++05,Gav++07}.

Strong lensing observations demonstrate the existence of DM halos
around individual massive early-type galaxies out to $z\sim1$ beyond
any reasonable doubt, both by themselves and in combination with other
techniques (for early-type galaxies with $\sigma \lesssim$ 200
kms$^{-1}$ the case is much less conclusive; future sample of low-mass
deflectors may be needed to clarify matters). One argument is that the
amount of mass inside the Einstein radius exceeds the stellar mass
M$_*$. This latter quantity can be constrained in many ways. Assuming
an IMF, stellar population synthesis (SPS) models applied to
photometric or spectroscopic data yield M$_*$ with an uncertainty of
0.1-0.2 dex. Alternatively, local dynamical studies of early-type
galaxies \citep{Ger++01,Cap++06} constrain the stellar mass-to-light
ratio at present time, which can then be evolved back in time either
using the measured evolution of the fundamental plane or other
measurement of the star formation history
\citep[e.g.,][]{Koc95,T+K04}.

A particularly powerful combination for detecting DM halos is to use
stellar kinematics of the lens galaxy to provide information on the
distribution of mass in the high surface brightness regions well
within the effective radius, and to use strong lensing to help remove
the mass-anisotropy degeneracy \citep[e.g.][]{T+K04,Bar++09}. A third
method relies on assuming scaling relations to analyze lenses across a
sample and reconstruct the mass density profile for the ensemble,
which turns out to be more extended than expected if mass followed
light and therefore consistent with DM \citep[][]{R+K05,Bol++08b}. A
fourth method exploits microlensing statistics to demonstrate that
point masses (i.e. stars) cannot contribute the totality of the
surface mass density at the location of the multiple images
\citep[e.g.][]{Poo++09}. A fifth method consists of measuring time
delays between multiple images, determining angular-diameter distances
from independent cosmographic probes to infer the behavior of the mass
density profile at the location of the multiple images
\citep{Koc++06a}.

\subsubsection{What is the relative spatial distribution of luminous and dark matter?}

The efficiency with which baryons condense inside halos to form stars,
and their effect on the underlying DM distribution, depend on the
interplay between cooling and heating (e.g. from star formation and
nuclear activity).  Lensing can help us understand these process by
providing precise measurements of the fraction of total mass in the
form of DM (\fdm) within a fixed projected radius, typically expressed
in terms of fraction of the effective radius \citep[e.g.][]{J+K07}.

\begin{figure}
\centerline{\psfig{file=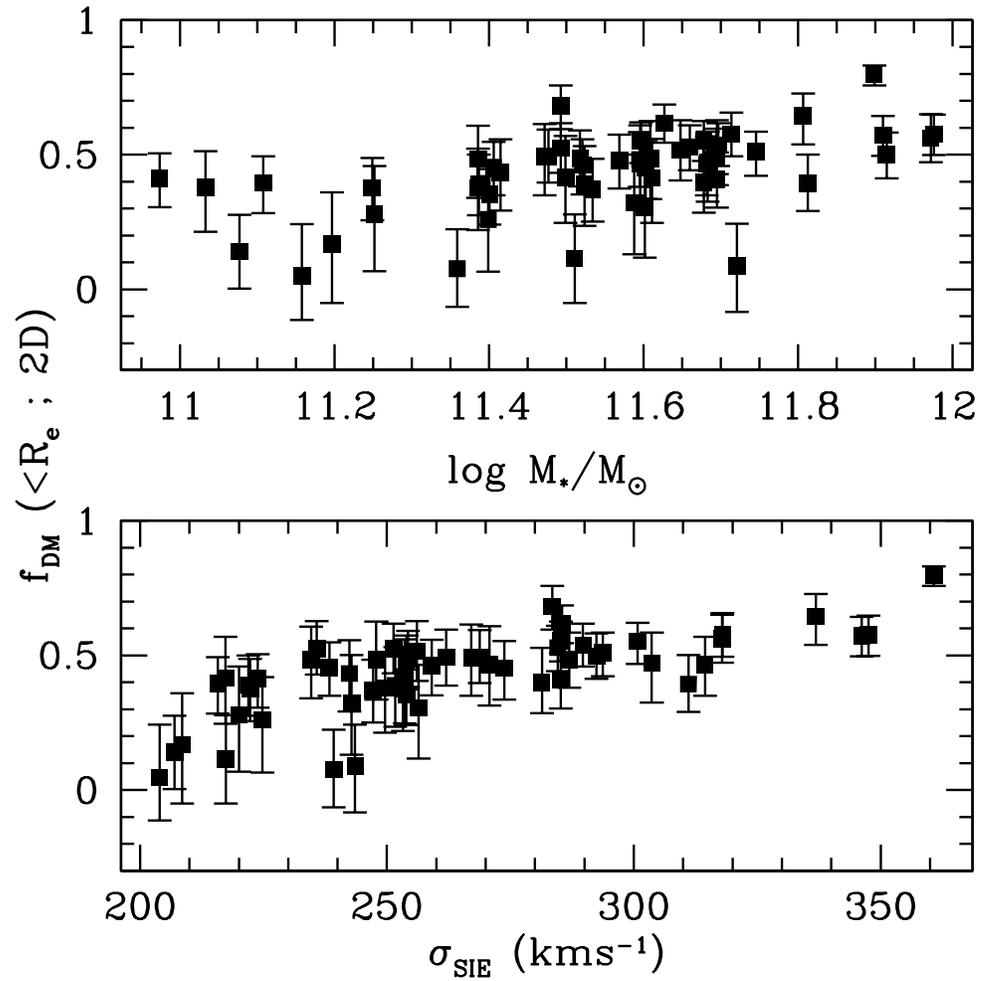,width=\textwidth}}
\caption{Dark matter fraction inside the cylinder of projected radius equal 
to the Einstein radius as inferred from stellar population synthesis
modeling of multicolor data and strong gravitational lensing analysis
of the SLACS sample \citep[data from][]{Aug++09a}.}
\label{fig:fdm}
\end{figure}

Observationally, \fdm\, is found to be non-negligible already at the
effective radius \citep[$25\pm6$
\%][]{Koo++06} and increasing towards larger radii \citep[$70\pm10$\%
at five effective radii][]{T+K04}.  Consistent results are obtained by
a number of independent non-lensing techniques
\citep[e.g.][]{Cap++06}. In addition, \fdm\, within a fixed fraction
of the effective radius is found to increase with galaxy stellar mass
and velocity dispersion.  For example, by comparing lensing masses
with those inferred from SPS modeling of multicolor data, \fdm\,
inside the cylinder of projected radius equal to the Einstein radius
increases from $\sim 25$\% to $\sim75$\% in the range of velocity
dispersion $\sigma$=200-350 \kms, or equivalently in the range of
stellar mass between 10$^{11}$ and 10$^{12}$ M$_{\odot}$
\citep[Figure~\ref{fig:fdm}]{Aug++09a}. 
These numbers are based on a \citet{Sal55} IMF and are consistent with
those inferred by local dynamical studies
\citep[e.g][]{Cap++06}. Adopting a \cite{Cha03} IMF changes the
overall normalization, but not the global trend \citep[][see, however,
Grillo et al.\ 2009 for a contrasting view]{Aug++09a}.

Strong lensing studies also explain the origin of the so-called
tilt of the fundamental plane \citep[e.g.,][hereafter FP]{CLR96},
the tight correlation between effective radius, effective surface
brightness and stellar velocity dispersion observed for early-type
galaxies. By introducing a dimensional mass variable M$_{\rm dim}
\equiv \sigma^2 {\rm R}_{\rm e} / {\rm G}$, the FP can be cast
in terms of an increasing effective mass-to-light ratio with effective
mass (the tilt'). Exploiting strong lensing, a somewhat tighter mass
plane \citep[MP][]{Bol++08b} relation can be obtained by replacing
surface brightness with total surface mass. The MP is not tilted,
implying that the tilt of the FP stems from an increase in
\fdm\, with mass, and not in a systematic change, e.g., of the virial
coefficient that connects M$_{\rm dim}$ to total mass.

\subsubsection{Mass density profiles and the bulge-halo conspiracy}

Another quantity of interest is the average logarithmic slope of the
three-dimensional total mass density profile $d \log \rho_{\rm tot} /
d
\log r \equiv - \gamma'$. An isothermal mass model has $\gamma'=2$. The
total mass density profile for a spherical model is often expressed in
terms of the equivalent circular velocity
\begin{equation} 
v_c \equiv \sqrt{\frac{GM(<r)}{r}},
\label{eq:rotcurve}
\end{equation}
which facilitates comparison with the literature on spiral galaxies
and on numerical simulations.  For a spherical power-law density
profile, $\gamma'$ is simply related to the slope of the rotation
curve by the relation $ d \log v_c / d \log r = (2-\gamma')/2$. For this
reason, an isothermal profile is sometimes referred to as a flat
rotation curve.

The basic result on this topic is that $\gamma'\approx 2$,
i.e. early-type lens galaxies have approximately isothermal mass
density profiles, or close-to-flat equivalent rotation curves. This
has been known since at least the early nineties, both on the basis of
lensing studies \citep[e.g.][]{Koc95} and on local kinematics
\citep[e.g.][and references therein]{B+S93,Ger++01}. However, in order
to understand the mass structure of galaxies with sufficient level of
precision to constrain formation models, we need to ask more detailed
questions. What is the average $\gamma'$ and its intrinsic scatter for
the overall population of early-type galaxies?  How does $\gamma'$
depend on the galactic radius or other global properties? Does it
depend on the environment, as expected if halos were tidally
truncated? Does $\gamma'$ evolve with redshift? In addition, as we
will see in \S~\ref{sec:cosmo}, determining the mass profiles of lens
galaxies to high accuracy is essential for many applications to
cosmography.
  
\begin{figure}
\centerline{\psfig{file=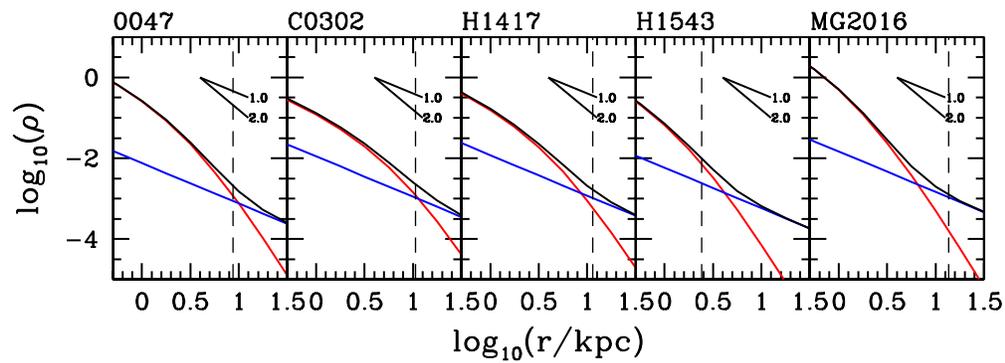,width=\textwidth}}
\caption{Mass density profiles of lens galaxies inferred from a strong
lensing and dynamical analysis \citep[Figure from][reproduced by
permission of the AAS]{T+K04}. In addition to the mass associated with
the stars (red line), the data require a more extended mass component,
identified as the dark matter halo (blue line). Although neither
component is a simple power-law, the total mas profile is close to
isothermal, i.e. $\gamma'=2$. The vertical dashed line identifies the
location of the Einstein radius.}
\label{fig:conspiracy}
\end{figure}

In the past few years, the large number of lenses discovered and the
high level of precision attainable with lensing has enabled
substantial breakthroughs. Joint lensing and dynamical studies of the
SLACS sample have shown that $\gamma'=2.08\pm0.02$ with an intrinsic
scatter of less than 10\% \citep{Koo++09}. This result is valid in the
sense of an average slope inside one effective radius or less, the
typical size of the Einstein radius of SLACS lenses. For higher
redshift deflectors, Einstein radii are typically larger than the
effective radius and reach out to 5 R$_{\rm e}$. Although the high
redshift samples with measured velocity dispersions are small, they
seem to suggest a somewhat larger intrinsic scatter around $\gamma'=2$
\citep{T+K04}. No significant dependency on galactic radius, global
galaxy parameter, or redshift has been found so far based on lensing
and dynamical analysis \citep{Koo++09}. The small scatter around
$\gamma'=2$ is remarkable, considering that neither the DM halo, nor
the stellar mass are well described by a simple power-law
profile. Nevertheless, the two components add up to an isothermal
profile (Fig.~\ref{fig:conspiracy}).  This effect is similar to the
disk-halo conspiracy responsible for the flat rotation curves of
spiral galaxies \citep{V+S86}, and it is therefore been dubbed the
'bulge-halo conspiracy'.  Detailed dynamical studies of the
twodimensional velocity field of deflector galaxies in conjunction
with strong gravitational lensing confirm this picture to higher
accuracy \citep{Bar++09}.

\begin{figure}
\centerline{\psfig{file=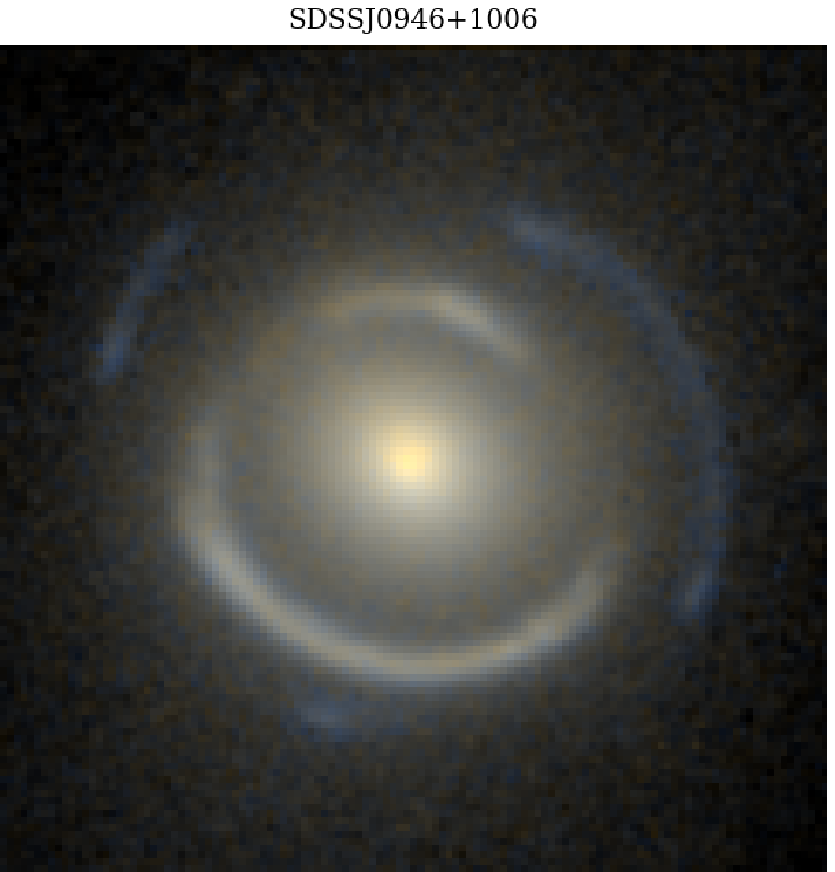,width=0.5\textwidth}
\psfig{file=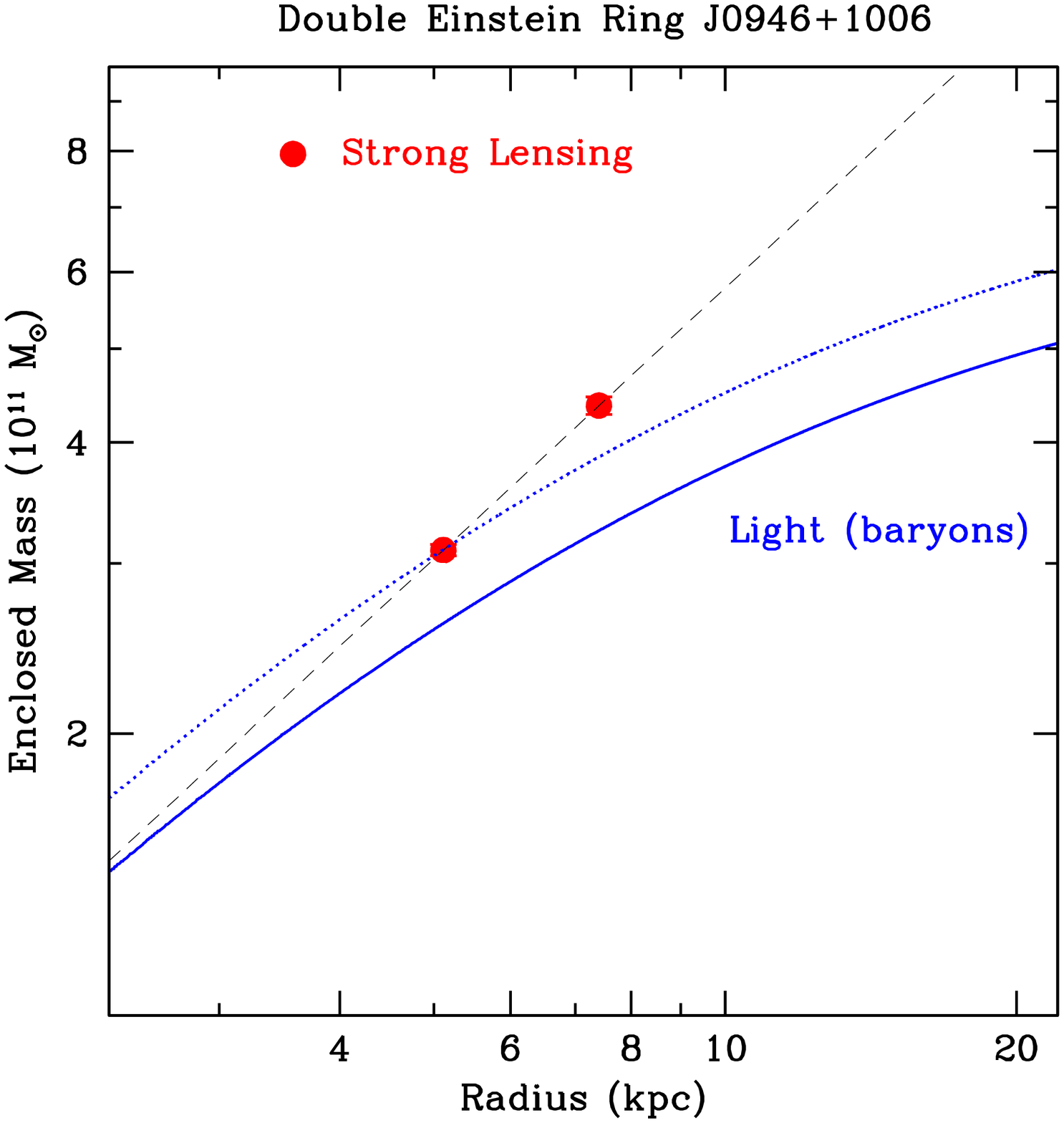,width=0.5\textwidth}}
\caption{Double Einstein ring compound lens SDSSJ0946+1006. {\bf
Left:} color composite HST image (Courtesy of M.~W.~Auger). Note the
foreground main deflector in the center, the bright ring formed by the
images of the intermediate galaxy, and the fainter ring formed by the
images of the background galaxy lensed by the two intervening
objects. {\bf Right:} Enclosed mass profile as inferred from the
Einstein radii of the two rings (red solid points - the error bars are
smaller than the points). The enclosed mass increases more steeply
with radius than the enclosed light (solid blue line; rescaled by the
best fit stellar mass-to-light ratio), indicating the presence of a
more extended dark matter component. Even a ``maximum bulge'' solution
(dotted blue line) cannot account for the mass at the outer Einstein
radius.}
\label{fig:double}
\end{figure}

Similar and consistent results can be obtained directly from
gravitational lens models, both for lensed sources covering a
significant radial range \citep[e.g.][]{D+W05} or when a gravitational
time delay has been measured and the cosmology is fixed by independent
measurements \citep{Koc++06a}. An interesting case is that of the
system SDSSJ0946+1006 where the presence of two multiply-imaged
sources at different redshifts constrains the projected mass density
slope to be $\gamma'=2.00\pm0.03$, based purely on lens modeling
(Figure~\ref{fig:double}). The lack of central images also constrains
the slope of the total density profile to be steep (e.g., $\gamma'=2$)
in the central regions of deflectors.  It should be noted that lensing
is mostly sensitive to the projected mass density slope at the
location of the images, rather than the average inside the
images. Therefore, a direct comparison with the lensing and dynamical
results is only valid to the extent that a pure power-law profile is a
good model for the data.

\subsubsection{Are dark matter density profiles universal?}

Cosmological numerical simulations predict that DM density profiles
should be almost universal in their form
\citep[][hereafter NFW]{NFW97}.
Simulated profiles are characterized by an inner slope $d\log
\rho_{DM} / d \log r = - \gamma \approx - 1$. At the scales of spiral
galaxies, low surface brightness galaxies, and clusters of galaxies,
it has been shown that in a number of systems the observed profiles
are shallower than predicted \citep[i.e. $\gamma<1$,
e.g.,][]{Sal++07,San++08}. The discrepancy suggests that either the
DM component or the effects of baryons on the underlying
halos are poorly understood.

In early-type galaxies the inner regions are completely dominated by
stellar mass, making them particularly interesting systems for
understanding the interplay between baryons and DM. Unfortunately, the
dominance of baryons also makes the measurement more challenging. A
joint lensing and dynamical analysis of 5 high-z lenses shows that
$\gamma$ is consistent with unity, albeit with large errors, and
shallower slopes cannot be excluded \citep{T+K04}. Improving the
measurement will require larger samples of objects with good quality
data and further constraints on the stellar mass-to-light ratio.

Alternatively, by imposing $\gamma=1$ one can infer an absolute
normalization of the stellar mass component, and thus constrain the
IMF of massive early-type galaxies to be close to Salpeter
\citep{Gri++09,Tre++10}. A joint lensing, dynamical, and stellar
population analysis of the SLACS sample shows that massive early-type
galaxies cannot have both a universal DM halo and universal IMF
\citep{Tre++10}: either the inner slope of the DM halo or the
normalization of the IMF have to increase with deflector velocity
dispersion.

\subsubsection{Implications for early-type galaxy formation}

Massive early-type galaxies are simple dynamical systems with simple
stellar populations. Yet, their formation and evolution is still far
from being well understood (for a comprehensive review see S.M.~Faber,
this volume). The standard CDM model postulates their formation via
major mergers, but this is hard to reconcile with their uniformly old
stellar populations -- unless there is some fine-tuned feedback
mechanism that prevents star formation in the high mass systems
\citep[see][for a recent review]{Ren06} -- and with the slow observed
evolution of their stellar mass function since $z\sim1$. Recently,
collisionless mergers not involving gas and star formation (and
therefore ``dry'') have become increasingly popular as a possible
mechanism of growth (Faber, this volume). Furthermore, dry mergers can
grow galaxies in size faster than in velocity dispersion. Therefore
they have been suggested as a possible mechanism for the evolution of
ultradense massive galaxies at high redshift into the more diffuse
ones found in the local universe \citep{vdW++09}.

Strong lensing studies give us some direct information on the
connection between baryons and DM, and therefore offer us new insights
into this problem. The (non-evolving) isothermality of the total mass
density profile requires an early dissipative phase, to steepen the
NFW profiles predicted in CDM-only simulations. Alternatively, an
initial collapse associated with incomplete violent relaxation could
have established the isothermality of the inner profiles. Either
phenomenon must have occurred well before $z\sim1$. After the initial
formation, further growth by dry mergers preserves the isothermal
profile and tightness of the mass plane
\citep{Koo++06,NTB09}. However, dry mergers do not preserve the tight
correlations between size and total mass and velocity dispersion and
total mass \citep{NTB09}. The observed tightness of the correlation
limits the growth by dry mergers to have been at most a factor of two
since $z\sim2$, unless there is a large degree of fine tuning between
orbital parameters of the merger and location in the
size-mass-velocity dispersion space. Therefore, it seems most likely
that the majority of the mass assembly must have occurred during the
initial dissipative phase associated with the dominant episode of star
formation.

The other main strong lensing result, i.e. the correlation between DM
fraction and velocity dispersion (stellar mass), provides us with
another piece of the puzzle. Dry mergers increase $f_{dm}$
\citep{NTB09}, thus creating part of the trend. However, dry-mergers
cannot explain the whole trend, which must be largely established
early-on through other means. A scenario where the time since major
initial collapse increases with present-day mass could explain the
trend in terms of the evolution of the density of the universe with
cosmic time \citep{Tho++09}. The correlation between present day mass
and epoch of major mass assembly could also help explain the
correlations between present day mass, age, and chemical composition
of the stellar populations \citep{Tre++05b}.

It should be noted that the conclusions above hold only for the most
massive early-type galaxies. At lower masses, evolution is certainly
more recent and other secular or environmentally driven mechanisms
could be responsible for forming early-type galaxies
\citep[e.g.][]{BTE07}.

\subsection{Luminous and dark matter in spiral galaxies}

Massive DM halos around local spiral galaxies are readily detected
from the gas kinematics at large radii \citep{V+S86}. The total
gravitational potential can be reconstructed accurately from the
observed velocity field. However, decomposing the total mass
distribution into its baryonic and dark components for individual
galaxies is still an unsolved problem, largely because the stellar
mass-to-light ratio is uncertain by a factor of $\sim$2-3 for young
and dusty stellar populations.  In the distant universe, the problem
is compounded by observational difficulties: HI becomes prohibitively
expensive to detect; optical rotation curves can be measured out to
$z\sim1$ but are limited by cosmological surface brightness dimming as
well as angular resolution.  One approach consists of assuming that
the baryonic component is maximally important, the so-called
maximum-disk ansatz \citep{V+S86}. However, it is not clear that disks
are indeed maximal. Indeed, submaximal disks seem to be suggested by a
variety of arguments \citep[e.g.][]{C+R99}, even though the unknown
IMF is a dominant source of uncertainty \citep{B+d01}. Understanding
the relative mass in disks and halos is critical to formulate and test
a robust theory of disk galaxy formation
\citep[e.g.][]{Dut++07}.

Gravitational lensing provides a new tool for luminous and dark matter
decomposition in spiral galaxies. Two factors make lensing
particularly useful in this respect. Firstly, it measure the total
projected mass within a cylinder. This can then be combined with the
enclosed mass in 3D inferred from disk kinematics to break the
disk-halo degeneracy by exploiting the different radial dependency of
the two components \citep[e.g.,][]{MFP97}. Secondly, gravitational
lensing provides azimuthal information which also helps pin down the
relative contribution of the two, especially if they are misaligned.

Strong lensing studies of spiral galaxies have shown encouraging
results, although the impact of the conclusions is limited by the
small size of current samples. For example, \citet{Tro++10} combined
lensing constraints, high resolution imaging data, and optical and
radio kinematics to decompose the mass profile of the Einstein Cross
lens galaxy into its bulge, disk, and halo components (see also van de
venn et al. 2010, submitted to ApJ). The mass-to-light ratio of the
bulge is very well constrained (M/L$_B=6.6\pm0.3$ in solar units). Due
to the unusually small Einstein radius of this system, the mass of the
disk is less well constrained, although it is clearly sub-maximal,
contributing $45\pm11$\% of rotational support at 2.2 scale lengths.
  
\begin{figure}
\centerline{\psfig{file=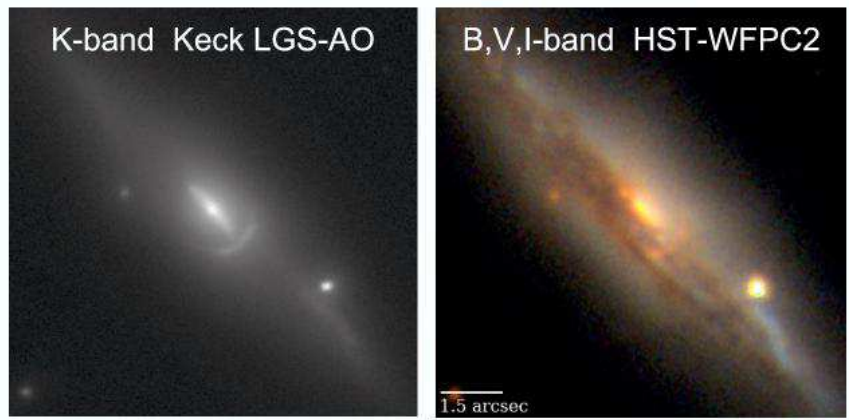,width=\textwidth}}
\caption{Example of edge-on spiral lens system ($z_d=0.063$) discovered
by the SWELLS Survey. The multiply-imaged source ($z_s=0.637$) is
visible in the optical HST discovery image and readily apparent in the
Keck near infrared image where the effects of dust are minimized. The
combined information at multiple wavelengths allows one to correct for
dust and infer the stellar mass of the disk (Image credits: A.~Dutton,
P.~Marshall, T.Treu).}
\label{fig:spiraldust}
\end{figure}

The situation is changing rapidly, due to progress in strong lensing
searches. SLACS discovered approximately 7 new bulge-dominated spiral
lenses and an ongoing search based on a similar-strategy (SWELLS;
HST-GO-11978) should find as many edge-on late type spirals. Dedicated
searches \citep[e.g.,][]{F++09,Mar++09} should discover tens of new
systems in the next few years. At variance with the smoothness of
early-type galaxies, the small-scale structure of the surface
brightness of the spiral lens due to dust and inhomogeneous stellar
populations complicates the identification and modeling of
multiply-imaged parts of the background source. High resolution
near-infrared images with adaptive optics or with HST and JWST,
coupled with multicolor optical data, or in the radio, will be
essential to make progress on this front (Fig.~\ref{fig:spiraldust}).

\section{Substructure in Galaxies and the ``Excess Subhalos'' Problem} 
\label{sec:sub}

\subsection{Background}

In the standard cosmology, DM halos host a hierarchy of
sub-halos, also known as DM substructure.  The number of
subhalos above a given mass scales approximately as the total mass of
the parent halo, and the logarithmic slope of the subhalo mass
function is approximately $dN/dM_{\rm sub} \propto M_{\rm
sub}^{-\alpha}$, with $\alpha=1.9\pm0.1$
\citep{Die++08,Spr++08}. Remarkably, the normalized distribution of
substructure depends very little on the overall scale of the halo,
therefore we would expect approximately the same abundance of
satellites around clusters and galaxies.

Although realistic simulations including baryons and non-gravitational
effects have yet to be performed at this scale, it is currently
believed that the statistical properties of the substructure inferred
from N-body simulations should be robust enough to allow for a direct
comparison with observations \citep[see, e.g., ][and references
therein]{Kra10}.  For these reasons such a comparison may provide one
of the most stringent and direct tests of the CDM paradigm at
subgalactic scales.

At variance with the results of simulations, the abundance of luminous
satellites observed around real clusters and galaxies are very
different. Whereas clusters of galaxies host thousand of galaxies
within their own DM halos, fewer satellites are generally
seen around galaxies. In particular, the mass function of the luminous
satellites of the Milky Way differs dramatically from that of the
subhalos of a typical simulated halo of comparable mass. At the high
mass end of the distribution (virial M$_{\rm sub}\sim 10^9 M_\odot$)
the observed number of satellites is comparable, or perhaps even
slightly larger, than expected. However, the mass function of the
halos of the observed satellites is found to be much shallower than
that predicted for subhalos, resulting in a dramatic shortfall at
lower masses, below 10$^8 M_{\odot}$. This discrepancy between theory
and observations has been known for over a decade
\citep{Kly++99,Moo++99}, and has not been solved by the revolutionary
discovery of low luminosity satellites of the Milky Way by SDSS, nor
by advances in numerical simulations. An up-do-date summary of the
current state of the problem is given by
\citet{Kra10}.

\subsection{Possible solutions}

There are two classes of possible explanations for the so-called
``excess sub-halos problem'' (or ``missing satellites problem'' if you
are a theorist). One possible explanation is that substructure exists,
but it is dark, i.e. subhalos do not form enough stars to be
detected. This explanation would imply that the conversion of baryons
into stars is inefficient for small halos. It is hard to explain this
inefficiency with the known mechanisms of supernovae feedback or the
effect of the UV ionizing background \citep{Kra10}. Alternatively, it
is possible that subhalos are not as abundant as predicted by
numerical simulations. This explanation would imply a major revision
of the standard CDM paradigm, either reducing the amplitude of
fluctuations on the scales of satellites, or changing the nature of DM
from cold to warm \citep{M+M07}. Either explanation has far reaching
implications. In order to be viable, the first explanation requires a
clear improvement in our understanding of galaxy formation. In its
most extreme version, the second explanation may require a re-thinking
of the paradigm.

Gravitational lensing provides a unique insight into this problem,
since it is arguably the only way to detect dark substructure, measure
its mass function, and compare it with the prediction of CDM numerical
simulations.  Even if advances in theories of galaxy formation could
explain the luminosity function of Milky Way satellites, there would
be still be a robust and falsifiable prediction of large numbers of
darker satellites to be tested.

If the mass function of sub-halos turns out to be different than that
predicted by simulations a major revision of the theory would be
required, possibly requiring warm DM, although it is not
clear than that would necessarily be a compatible with all other
constraints \citep[see][and references therein]{Kra10}.

\subsection{Flux ratio anomalies}

The most striking and easiest to detect lensing effect of substructure
is the perturbation of the magnification pattern. Since magnification
depends on the second derivative of the potential, a small local
perturbation can introduce dramatic differences in the observed
surface brightness of the lensed source, without altering
significantly the overall geometry of the system. For point sources,
the presence of substructure results in ratios of the fluxes of
multiple images that are significantly different than what would be
predicted by a smooth macro model. This effect is often referred to as
the anomalous flux ratios phenomenon, and has been used to infer the
presence of substructure in lens galaxies
\citep{M+S98,Chi02,D+K02,Bra++02,M+Z02}. In an influential paper,
\citet{D+K02} analyzed radio data for a sample of seven quadruply
imaged sources, and reported the detection of a surface mass fraction
in the form of substructure between 0.6\% and 7\%.  This observed
fraction appears to be even higher than the mass fraction in
substructure at the Einstein radius predicted by simulations
\citep{Mao++04,Xu++09}.

Substantial efforts have been devoted to investigate whether
satellite-size halos are the most likely explanation of the observed
flux ratio anomalies. Indeed, flux ratio anomalies could also arise
from other effects such as microlensing -- if the source is
sufficiently compact -- or a non-uniform interstellar medium which
could variously affect light propagating along different paths.
However, both contaminants are wavelength dependent, while flux ratio
anomalies due to the substructure are achromatic. Therefore,
observations at multiple wavelengths, especially radio, narrow
emission lines, and mid-infrared, can be used to show that the the
anomalous flux ratios are effectively due to substructures on scales
much larger than stars \citep[e.g.,][]{AJB00,K+D04,M+M03}. Angular
structure in the macro model has been suggested as a possible cause
for flux ratio anomalies \citep{E+W03}. However, in the cases when
enough azimuthal information is available it has been shown that the
angular structure of lens galaxies is fairly simple and well
approximated by an ellipse
\citep{Yoo++06}.  Elegant arguments based on the local curvature of
the time-delay surface near the multiple images have also been used to
show that anomalous flux ratios are indeed due to mass substructure
\citep{K+D04,Che09}. A final source of concern is potential
contamination from substructure along the line of sight, which could
mimic the effects of true galactic satellites
\citep{CKK03,Che09}. Line-of-sight contamination is most likely
not the main cause of the anomalies observed so far. However, it is
clear that line-of-sight contamination needs to be better understood
and quantified in order to extract the maximum amount of information
from this powerful tool.

An important question is whether the detected substructure is dark or
luminous. In some cases \citep[e.g.][]{Koo++02,McK++07,MKA09} it has
been shown that mass associated with luminous satellites can explain
the observed anomalies. Whether luminous substructure can explain all
the known anomalies is still a matter of debate
\citep{Che09}. On a case by case basis, the role of luminous
satellites is difficult to quantify because they are hard to detect in
the vicinity of the bright lensed quasars, where they would be most
effective in introducing anomalies. In addition to high resolution HST
or adaptive optics images, an accurate determination of the luminosity
function and spatial distribution of luminous satellites of
(non-lensing) massive galaxies may be a way to make progress. The
challenge is to collect large enough samples of non-lenses while
carefully matching the selection process of the sample of lenses. It
is important to stress that the detection of optical counterparts does
not undermine the quest for substructure using gravitational
lensing. Measuring the mass function of satellites - whether they are
visible or not - is essential to test the CDM paradigm. Comparing the
satellite mass function with their luminosity function will only help
in answering some of the questions related to the mechanisms which
regulate star formation.

The detection of substructure via anomalous flux ratios is an example
of the power of gravitational lensing in measuring the distribution of
mass in the universe. However, the strong lensing studies to date
suffer from two fundamental limitations, which need to be overcome in
order to make progress. The first limitation is poor and uncertain
statistics. Not only is the number of systems that can be used to
study anomalous flux ratios tiny, but the selection function is poorly
characterized. Therefore, the uncertainties are large and the results
could be biased. The second major limitation is the limited mass
sensitivity achieved so far, which is only sufficient to probe the
upper end of the mass function of subhalos.

Major improvements on both aspects are underway and significant
progress is possible in the next few years. One key factor is the
increase in the number of known lenses, discovered with a well defined
selection algorithm, coupled with the increased capability for
follow-up. In the next decade, we may expect tens of thousands of
lenses to be discovered by radio and optical surveys
(\S~\ref{sec:future}). The other key factor is the development of
advanced techniques to be applied to high resolution data to probe
further down the mass function of subhalos, discussed next.

\subsection{Astrometric and time-delay anomalies}

Flux ratio anomalies is only one way to detect substructure. Subhalos
affect all lensing observables, including deflection angles and time
delays, and can therefore be detected as corresponding perturbations
with respect to the predictions of a smooth model.  Although these are
more subtle effects, they have been shown to be sufficiently large to
be used to detect substructure \cite[e.g.][]{Che++07,K+M09}.
Galaxy-galaxy lenses where the multiple images form an almost complete
Einstein ring and are observed with high signal-to-noise ratio can
detect individual substructures with masses as low as $\sim$10$^8$
M$_\odot$ \citep{Koo05,V+K09a}. Recent calculations by \citet{V+K09}
indicate that current samples of galaxy-galaxy lens systems such as
SLACS can detect subhalo mass fractions as low as 0.5\%, assuming the
slope of the mass function is well known from simulations. A sample of
200 Einstein rings with data of comparable quality to HST should be
sufficient to start constraining the slope of the mass function as
well. The sensitivity will be further enhanced with advances in
resolution expected from future radio telescopes and the next
generation of adaptive optics systems on large and extremely large
telescopes. Furthermore, anomalous flux ratios, astrometric
perturbations, and time delay anomalies depend on different moments of
the satellite mass function \citep{Kee09}. Therefore, a combination of
techniques can help constrain both the slope and the normalization of
the substructure mass function.

\section{Cosmography}
\label{sec:cosmo}

Cosmography is the measurement of the parameters that characterize the
geometry, content, and kinematics of the universe. Much progress has
been achieved in recent years \citep[e.g.,][]{Kom++09}, heralded as the
era of precision cosmology. However, some of the fundamental
parameters need to be measured even more accurately if one wants to
discriminate between competing theories. For example, the equation of
state of dark energy $w$ and its evolution with cosmic time are
essential ingredients to understand the nature of this mysterious
phenomenon.

Strong lensing is a powerful cosmographic probe, as it depends on
cosmological parameters in two ways. Firstly, the time delay equation
(and the lens equation) contain ratios of angular diameter
distances. Therefore, within the context of a model for the lensing
potential, measurements of time delays or mass act as standard rods,
in a similar manner as the acoustic peaks of the power spectrum of the
cosmic microwave background. Cosmography based on this concept is
described in Sections~\ref{ssec:timedelay},
\ref{ssec:standard}, and~\ref{ssec:compound}. Secondly, the optical
depth for strong lensing depends on the number and redshift
distribution of deflectors and therefore on the growth of structure
and on the relation between redshift and comoving volume. Thus, given
a model for the lensing cross section, and a model for the evolution
of the population of deflectors, one can do cosmography from lens
statistics. This approach is described in~\S~\ref{ssec:statistics}.

\subsection{Time delays}
\label{ssec:timedelay}

\begin{figure}
\centerline{\psfig{file=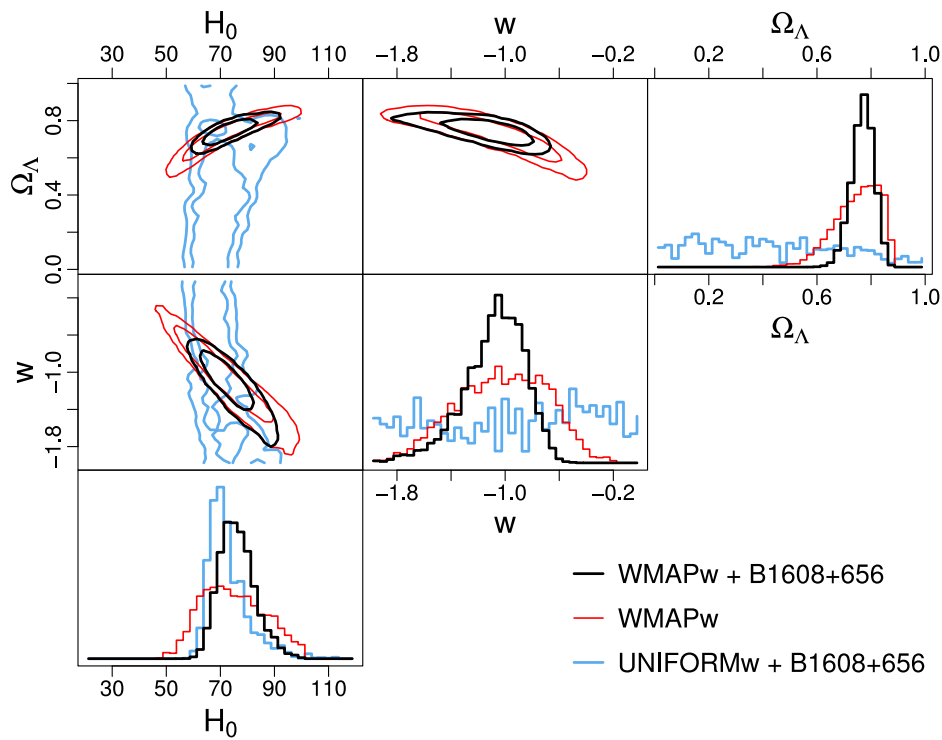,width=\textwidth,angle=270}}
\caption{Illustration of cosmography with gravitational time
delays. The panels show two and one dimensional posterior probability
distribution functions for H$_0$, $w$, and $\Omega_\Lambda$, assuming
flatness. Red lines indicate limits from cosmic microwave background,
blue lines represent limits obtained from a single gravitational lens
system with measured time delays (B1608+656), while black lines
represent the joint constraints. Note how the constraints from
time-delays are almost vertical in H$_0$ and therefore help break the
degeneracy between $w$ and H$_0$ in the CMB data. Lensing constraints
in the $w-\Omega_\Lambda$ are broad and therefore not shown for
clarity (Figure courtesy of S.Suyu; Figure from Suyu et al. 2010) .}
\label{fig:H0w}
\end{figure}

Consider a galaxy lensing a time-variable source like a quasar or a
supernova. Under the thin lens approximation, multiple images will be
observed to vary with a delay which depends on the gravitational
potential as well on a ratio of angular diameter distances
(Equation~\ref{eq:fermat}). The ratio of angular diameter distances is
mostly sensitive to the Hubble Constant H$_0$ (hereafter $h$ in units
of 100 \kms Mpc$^{-1}$). However, time delays also contain
non-negligible information about other cosmological parameters,
especially if one considers a sample of deflectors and sources
spanning a range of redshifts \citep[e.g.][]{C+M09}. Therefore,
although it is convenient to think in terms of the Hubble constant as
the primary parameter, time-delays provide constraints in the
multidimensional cosmological parameter space. When combined with
other cosmology probes like the CMB power spectrum, time-delays are
very effective at breaking degeneracies such as that between H$_0$ and
$w$ (Figure~\ref{fig:H0w}).

From a practical point of view, cosmography with time-delays can be
broken into two separate problems: measuring time delays and modeling
the lensing potential, including matter along the line of
sight. Uncertainties in these two terms dominate the error budget and
they are independent. Therefore, in order to measure H$_0$ to 1\%
accuracy from one lens system one needs to know both quantities with
sub percent accuracy. Or, for a sample of N lenses, one needs unbiased
measurements with approximately half $\sqrt N$\% uncertainty on both
quantities.

\subsubsection{Measuring time delays}

Measuring time delays requires properly sampled light curves of
duration significantly longer than the time-delay between multiple
images.  Once an approximate time-delay is known, the measurement can
generally be refined by adapting the monitoring strategy, e.g. with
dense sampling triggered after an event on the leading image. Typical
time delays for galaxy lens systems are in the range weeks to months
(with tails on both ends out to hours to years) and minimum detectable
amplitudes from the ground are of order $\sim$5\%, limited by
photometric accuracy for crowded sources and microlensing (see
\S~\ref{ssec:microlensing}). Thus, accurate time-delays typically
require several seasons of dedicated monitoring effort.

After the first ``heroic'' campaigns of the nineties and early 2000
\citep[see][for a review]{SKW06}, which yielded of order 10
time-delays, several groups are now trying to take this effort to the
next level with the help of queue mode scheduling and robotic
telescopes. A recent summary of published time-delay measurements is
given by \citet{Jac07}. Two new time-delays have been published since
then \citep[J1206+4332 and J2033-4723;][]{Par++09,Vui++08}. Taking the
published time-delay uncertainties at face value, the present sample
contributes to the error budget on H$_0$ a little less than 1\%.  As I
will discuss in \S~\ref{sec:future}, time-domain astronomy is a
rapidly growing field and it is likely that many of the logistical
problems faced by time-delay hunters so far will be solved in the next
decade.

\subsubsection{Determination of the lensing potential}

We now turn to errors associated with the local lensing potential,
under the single screen approximation (matter along the line of sight
and associated uncertainties will be described
in~\S~\ref{ssec:masssheet}). At fixed image configuration, time-delays
depend to first order on the effective slope of the mass distribution
in the annular region between the multiple-images
\citep[see ][and references therein for
discussion]{SKW06,S+W06}. For generic power-law models, at fixed
lensing observables, the inferred H$_0$ scales as
H$_0(\gamma')\approx(\gamma'-1)H_0(\gamma'=2)$. For many systems,
especially doubly imaged point sources, the lensing potential is
highly uncertain and dominates the error budget. Unaccounted
uncertainties in the mass model are the main culprits for the reported
discrepancies between time-delay determinations of H$_0$ as large as
$\sim$30\% \citep[e.g.][]{T+K02b}.

It is clear that some additional information is needed to bring the
error budget on the lens modeling in line with that from time-delays.
One approach consists of asserting some prior knowledge of the mass
distribution in the deflectors and applying it to the analysis of a
sample of systems. Since the effective slope is poorly constrained by
lens data for point-like sources without additional information, the
results depend critically on the prior. Following this approach,
\citet{Ogu07} modeled 16 systems with power-law models assuming a
Gaussian prior on $\gamma'$ centered on 2 and width 0.15 obtaining
$h=0.68\pm0.06\pm0.08$ (the large systematic error attempts to reflect
the large dispersion from system to system; however, it may also be
due to the inclusion of systems with questionable redshift, time-delay
or embedded in a complex cluster potential, which carries substantial
additional modelling uncertainties). The prior on $\gamma$ is
plausible but not strictly justified, since there are no independent
measurements for the sample. For example, just changing the mean of
the prior to $\langle
\gamma'\rangle=2.085^{+0.025}_{-0.018}\pm0.1 $ as found for the
SLACS sample, would increase the estimate of H$_0$ by 8\%, with an
additional systematic uncertainty of 10\%. A very similar approach is
that by \citet{Col08}, who imposes geometric priors to his pixelized
mass reconstructions and obtains $h=0.71^{+0.06}_{-0.08}$ from 11
systems. Although it would be useful to draw samples from the
\citet{Col08} prior and measure the effective distribution of
$\gamma'$, it appears that his smoothness and steepness constraints
create a distribution of effective slopes similar to that of
\citet{Ogu07}, explaining the agreement. These results are
encouraging. However, they illustrate the challenge of reaching 1\%
accuracy using this methodology. One needs to have sufficient external
knowledge of the distribution of mass in the sample of galaxies with
measured time-delays to construct a sufficiently accurate prior.

A more promising approach is to extract additional information for the
very systems with measured time delays using ancillary data in
addition to those available for the multiply-imaged point sources. In
Bayesian terms, this means making the likelihood more constraining so
as to reduce the relative importance of the prior. Following this
approach, \citet{WBB04} modeled the extended radio structure around
the lensed quasar in B0218+357 to infer $\gamma'=1.96\pm0.02$ and
$h=0.78\pm0.03$. \citet{Koo++03} modeled B1608+656 using the measured
stellar velocity dispersion and the HST images of the lensed host
galaxy to measure $\gamma'$ and infer $h=0.75^{+0.07}_{-0.06}\pm0.03$,
fixing $\Omega_m=0.3$ and $\Omega_\Lambda=0.7$, and negelecting
uncertainties due to the mass-sheet degeneracy, discussed in the next
section. A recent analysis of improved Keck and HST data of B1608+656
by \citet{Suy++10} using more general pixellated models for the
potential and the source, infers $h=0.706\pm0.031$, for the same
cosmology as \citet{Koo++03}, including uncertainties related to the
mass-sheet degeneracy. This results shows that modeling errors can be
reduced to a few percent per lens system, if sufficient observational
constraints are available. If the other cosmological parameters are
allowed to vary, one obtains the constraints shown in
Figure~\ref{fig:H0w}. The information from time delays is particularly
powerful when combined to the WMAP5 results \citep{Kom++09}, improving
from $h=0.74^{0.15}_{-0.14}$ and $w=-1.06^{+0.41}_{-0.42}$ to
$h=0.697^{+0.049}_{-0.05}$ and $w=-0.94^{+0.17}_{-0.19}$, for a flat
cosmology. The results from a single lens are comparable with those
from the local distance ladder method \citep[$h=0.742\pm0.036$ and
$w=-1.12\pm0.12$ in combination with WMAP5;][]{Rie++09} in terms of
precision, although they are based on completely different physics and
assumptions, and subject to different systematic errors.

\subsubsection{Mass along the line of sight and the mass-sheet degeneracy}
\label{ssec:masssheet}

The final and perhaps limiting factor at this point is the uncertainty
due to the unknown distribution of mass along the line of sight,
i.e. deviations from the single screen approximation. On the one hand
massive galaxies are typically found in groups. Group members and the
common group halo contribute additional shear and convergence at the
location of the main deflector. On the other hand, the ``cone''
between us the observer and source may be over or underdense, thus
perturbing the time-delays with respect to those expected in a
perfectly smooth and isotropic universe. Both effects can be thought
to first order as equivalent to adding an external convergence
$\kappa_{\rm ext}$ at the location of the deflector (which can be
negative if the line of sight is underdense).

Due to the mass-sheet degeneracy, $\kappa_{\rm ext}$ is undetectable
from dimensionless lensing observables. However, if we ignored its
presence and make the standard assumption of vanishing convergence
away from the lens to break the mass-sheet degeneracy, we would infer
a biased value of H$_0$ by a factor $1/(1-\kappa_{\rm ext})$
\citep[e.g.][]{SKW06}.

Independent measurements of mass, such as stellar velocity dispersion,
help break the degeneracy because they constrain the local mass
distribution. An unknown $\kappa_{\rm ext}$ leads to an overestimate
of the lensing mass, and therefore alters the inferred $\gamma'$ from
comparison with kinematics, counterbalancing the effects on H$_0$, but
not exactly. Measurements of the local environment
\citep[e.g.,][]{Mom++06,Fas++06b,Aug++07} also help, although the limiting
factor is the precision with which mass can be associated with visible
tracers.  A third approach consists of inferring the distribution of
effective $\kappa_{\rm ext}$ from high resolution numerical
simulations \citep{Hil++07}. The challenges of this third approach are
producing realistic simulations at kpc scales relevant for strong
lensing and understanding the selection function of the observed
samples well enough to select simulated samples in the same way. In
the case of B1608+656, the total uncertainty can be brought to 5\%
using a combination of the three approaches \citep{Suy++10}. Analyzing
a number of systems in similar detail will help uncover whether there
are any residual significant biases.

\subsection{Lenses as standard masses} 
\label{ssec:standard}

Lensing studies indicate that the ratio f$_{\rm SIE}$ between stellar
velocity dispersion measured within a standard spectroscopic aperture
and the normalization of the best fit SIE model ${\sigma_{\rm SIE}}$
is close to unity \citep[1.019$\pm$0.08 for the SLACS sample for a
concordance cosmology][]{Bol++08b}, consistent with our general
understanding of the mass distribution of early-type galaxies in the
local Universe. If f$_{\rm SIE}$ is known sufficiently well --
independent of cosmology -- lens galaxies could effectively be used as
standard masses plugging measurements of Einstein radius and stellar
velocity dispersion into the SIE version of Eq.~\ref{eq:thetaSIS}
\citep{GLB08}. Note that H$_0$ cancels out in the ratio of angular 
diameter distances. Unfortunately, our current understanding of the
mass-structure of deflectors and of the distribution of matter along
the line of sight is not sufficient for accurate cosmography
\citep{SBR10}. In some sense, the situation is similar to that of time-delay
cosmography, and similar methodologies could be applied to overcome
the limitations. The advantage of this method over time-delays is that
it can be applied to any lens regardless of the presence of a variable
source. The disadvantage is that the sensitivity of the angular
diameter distance ratio on cosmological parameters is weak.

\subsection{Compound lenses}
\label{ssec:compound}

The Einstein radius of a gravitational lens depends on the mass
enclosed and on ratios of angular diameter distances. For systems with
multiple sets of multiple images, such as SDSSJ0946+1006
(Figure~\ref{fig:double}), one can solve for both the mass
distribution and cosmography, provided that enough information is
available to constrain the distribution of mass in the region between
the Einstein rings. An additional complication is given by the mass
associated with the inner ring, which acts as an extra deflector,
making these systems compound lenses for the background source
responsible for the outer ring. \citet{Gav++08} calculate that a
sample of 50 systems like SDSSJ0946+1006 -- expected for future large
lens surveys -- should constrain the equation of state of dark energy
$w$ to about 10\% precision. As in the cases discussed above, the
issue is whether systematics associated with modeling the deflector
itself or the structure along the line of sight can be controlled with
sufficient accuracy. High quality spatially resolved kinematic
information should help constrain the mass model of the main
foreground deflector and of the inner ring.

\subsection{Lens statistics}
\label{ssec:statistics}

For a given source population, the fraction of strongly lensed systems
(i.e. the optical depth) depends on the cross section of the
deflectors and on the abundance of deflectors. Thus, measuring the
abundance of strongly lensed systems constrains the intervening cosmic
volume. This is the essence of lens statistics as a tool for
cosmography, although quantities such as the distribution of Einstein
radii and source redshifts also contain cosmographic information Note
that lens-driven surveys are not nearly as sensitive as source-driven
surveys \citep[see][and references therein for a theoretical
description]{SKW06}.

The state of the art of this cosmographic application is the analysis
of 11 CLASS and 16 SQLS samples \citep{Cha07,Ogu++08}, which yield
rather weak bounds on cosmological parameters \citep[e.g.,
$w=-1.1\pm0.6^{+0.3}_{-0.5}$][]{Ogu++08}.  Even though precision can
certainly be improved by increasing sample size, the ultimate limit is
set by systematic uncertainties. Accurate cosmography from strong
lensing statistics requires accurate knowledge of: i) the
mass-structure and shape of deflectors to compute cross-sections; ii)
the contribution to the cross-section from large scale structures;
iii) the number density of deflectors; iv) the source luminosity
function; v) the survey selection function.  Quantities i to iv need
to be known as a function of redshift. In conclusion, lens statistics
poses three additional challenges (iii-v) over those in common with
other cosmographic applications.

\subsection{Is lensing competitive?}

The studies mentioned in this section show that cosmography with
strong-lensing gives results in agreement with independent probes,
reinforcing the so-called concordance cosmology. However, the ultimate
test for a method is when it breaks new ground in terms of precision,
and the result is then confirmed independently. In my view,
time-delays are the cosmographic application that stands the best
chance of doing this for three reasons. First, two out of three major
problems (time delay measurement and local mass model) have been
solved and progress on the third (external convergence) is being
made. Secondly, the inferred constraints are well suited to break
degeneracies inherent to other methods such as the CMB power
spectrum. Thirdly, time-delays can be measured for a number of lenses
using relatively small ground based telescopes or will come for free
from future synoptic telescopes.  Fourthly, the method is completely
independent of the local distance ladder method and therefore provides
a valuable independent test on its systematic uncertainties (like
calibration and metallicity dependency of the cepheid-luminosity
relation). Lenses as standard masses and compound lenses seem to be
valuable cosmographic tools if they can be applied efficiently with
limited observational resources, perhaps ``piggy-backing'' on other
studies (\S~\ref{sec:future}).

What is certainly very exciting and unique is the ``inverse''
application of cosmographic applications: learn about galaxy structure
and evolution, on the basis of accurate cosmography from other
probes. As mentioned in \S~\ref{sec:mass}, time-delays, the
combination of lensing and dynamics, and compound lenses have all been
demonstrated to provide unique insights into the structure of distant
galaxies, which cannot be obtained in any other way. This is also true
for lens statistics, which can be used to determine the growth of the
galaxy mass function in a unique way, once the mass structure of each
galaxy is understood from the other means discussed above
\citep{Cha07,Mit++05}.

\section{Lenses as Cosmic Telescopes}
\label{sec:telescopes}

In a typical galaxy-scale strong lens system, the background source is
magnified by an order of magnitude. Exploiting this effect, lensed
galaxies at intermediate and high redshift can be studied with the
same level of detail as non-lensed galaxies in the local universe
(\S~\ref{ssec:small}). Furthermore, the host galaxies of bright active
galactic nuclei are ``stretched away'' from the wings of the point
spread function, enabling precise measurements of their luminosity and
size, and ultimately of the cosmic evolution of the relation between
host galaxy and central black hole (\S~\ref{ssec:agn}). Finally,
microlensing by stars provides us with unique spatial information on
the scale of the accretion disk, which is orders of magnitudes smaller
than anything that can be resolved from the ground at any wavelength
(\S~\ref{ssec:microlensing}).

\subsection{Small and faint galaxies}
\label{ssec:small}

The resolution of HST and the sensitivity of radio interferometers
mean that we know very little about the distant ($z>>0.1$) universe on
scales below $\sim 1$ kpc. Indeed, even in the nearby universe
($z\sim0.1$), large ground based surveys such as SDSS do not provide
much sub-kpc scale information. Yet, we know from the local volume
that small and faint galaxies are an essential ingredient of the
universe, acting as building blocks of more massive systems. Only with
the aid of gravitational lensing we can resolve sub-kpc scales and
determine the morphology and size \citep{Mar++07}, and kinematics of
small galaxies as well as trace the location of star formation and the
pattern of chemical abundances \citep{Sta++08,Rie++08}.  Furthermore,
flux magnification enables detailed spectroscopic studies that would
be prohibitive in the absence of lensing \citep{Sta++08}.  These pilot
studies show that intrinsic properties can be robustly recovered via
lens modeling. The rapid increase in the number of known lenses should
soon provide the large statistical samples needed for high impact
studies.

\subsection{Host galaxies of lensed active nuclei}
\label{ssec:agn}

In the local universe, massive galaxies are found to harbor central
supermassive black holes. Remarkably, the mass of the black hole
correlates with kpc-scale properties of the host bulge, such as
velocity dispersion, luminosity and stellar mass
\citep[e.g.][]{Gul++09}. This family of correlations has been
interpreted as evidence that black hole growth and energy feedback
from active galactic nuclei play an important role in galaxy formation
and evolution \citep[e.g.][]{HMT09}. However, the physics of the
interaction as well as the relative timing of galaxy formation and
black hole growth are poorly understood. Although the local relations
are an important constraint, observing their cosmic evolution is
necessary to answer some fundamental questions. Are the local
relations only the end-point of evolution, or are they established
early-on? Which comes first, the black hole or the host bulge?

\begin{figure}
\centerline{\psfig{file=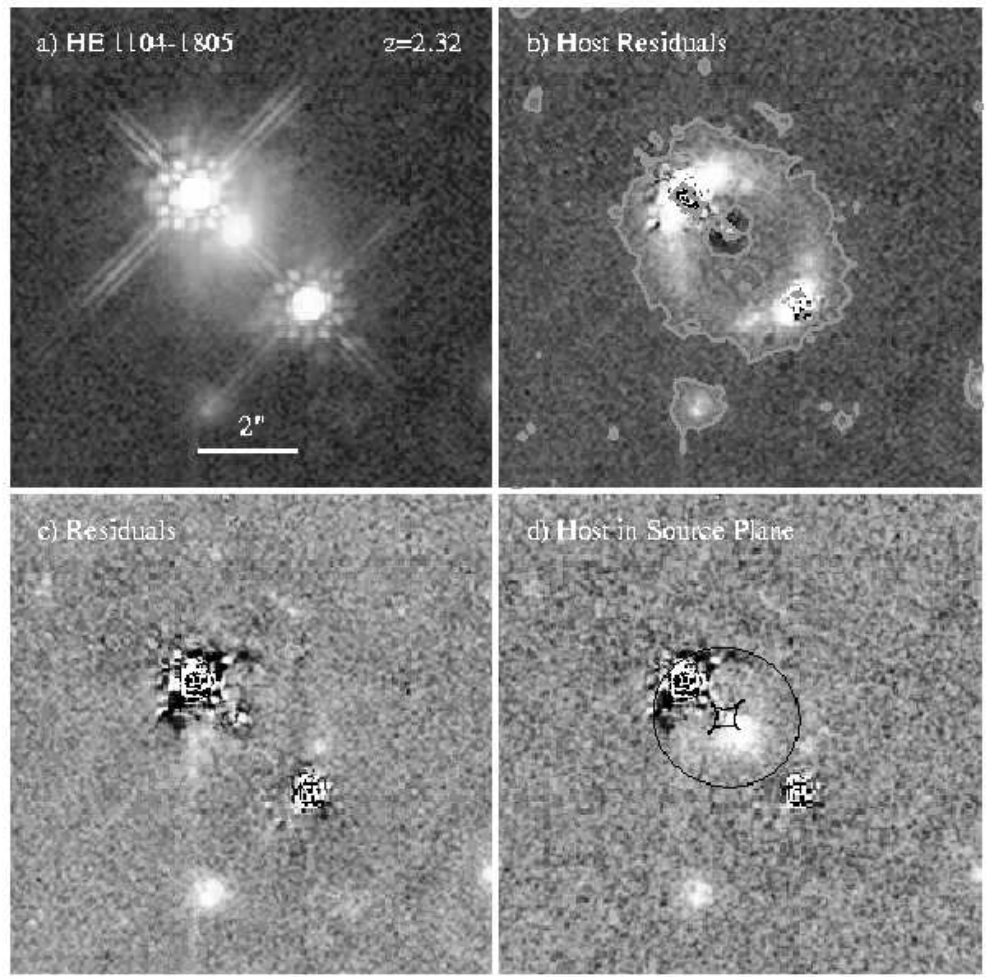,width=\textwidth}}
\caption{Illustration of gravitational lenses as cosmic
telescopes. Two-image lens system (HE 1104-1805) of a $z_s = 2.32$
quasar produced by a $z_d = 0.73$ foreground galaxy. Panel a) shows
the original data, panel b) shows the lensed host galaxy found after
subtracting the deflector and quasar components of the best-fitting
photometric model, panel c) shows the residuals from that photometric
model, and panel d) shows what the unlensed host galaxy would look
like in a similar exposure after perfectly subtracting the flux from
the quasar.  The curves shown superposed on the model of the host
galaxy are the lensing caustics. \citep[Figure from][reproduced by
permission of the AAS]{Pen++06qsob}}
\label{fig:PENG}
\end{figure}

It is challenging to answer these questions observationally. Direct
dynamical black hole mass measurements can only be done in the very
local universe. At intermediate and high-z redshift, one needs to rely
on indirect methods such as the empirically calibrated relation with
continuum luminosity and line width observed for type-1 active
galactic nuclei (AGN). However, the presence of bright luminous point
sources hampers the study of the host galaxy
\citep{Tre++07,Jah++09}. Strong lensing helps by stretching the host
galaxy of distant lensed quasars primarily along the tangential
direction (Figure~\ref{fig:PENG}). Of course, the quasar is also
magnified, but one generally wins because the surface brightness of
the point spread function falls off more rapidly than linearly.  Using
this method, \citet{Pen++06qsob} showed that the bulges of host
galaxies of distant quasars are more luminous than expected based on
the local relation, consistent with a scenario where bulge formation
predates black hole growth, at least for some objects. Similar results
have been found for non-lensed AGN \citep{Tre++07}. However, without
the aid of lensing, studies have to be limited to lower redshifts and
lower luminosity AGNs.

\subsection{Structure of active galactic nuclei}
\label{ssec:microlensing}

Understanding the physics of accretion disks and the regions
surrounding supermassive black holes is essential to explain the AGN
phenomenon with all its implications for galaxy formation and
evolution. However, the scales involved are extremely small by
astronomical standards (for a typical 10$^{9}$\, M$_\odot$ black hole,
the Schwarzschild radius is $\approx3\cdot10^{14}$\, cm, the broad line
region is $\sim 10^{17-18}$\, cm), and therefore impossible to resolve
with conventional imaging techniques. 

Microlensing is perhaps the only tool capable of probing the small
scales of the accretion disk. The Einstein radius of a star of mass
M$_s$ (Fig.~\ref{fig:lens}), corresponds to approximately
$4\cdot10^{16} \sqrt{{\rm M_s}/{\rm M}_{\odot}}$cm$\approx
0.01\sqrt{{\rm M_s}/{\rm M}_{\odot}}$pc when projected at the redshift
of a typical lensed quasar \citep[$z_d=0.5$; $z_s=2$][]{SKW06}.  The
inner parts of the accretion disk will be smaller than this scale and
therefore subject to microlensing, while the broad line region and the
outer dusty torus should be largely unaffected. The characteristic
timescale for variation is given by the microlensing caustic crossing
time, typically of order years, although it can be shorter for special
redshift combinations such as that of Q2237+030 \citep{SKW06}.

Based on this principle, one can infer the characteristic size of the
accretion disk as a function of wavelength. Long-light curves -- where
the gravitational time delay between multiple images can also be
determined -- provide the most stringent limits \citep{Koc04}, but
interesting information can also be obtained from single epoch data on
a statistical basis \citep[e.g.][]{BWW07,Poo++09}.

The inferred absolute size of the accretion disk can be known up to a
factor of order unity, which depends on $\langle M_s \rangle$ and on
the relative transverse speeds between the stars, the deflector, and
the source. However, the slope of the relation between accretion disk
temperature and size is independent of that factor and can thus be
determined more precisely. Current results indicate that the accretion
disk is approximately the size expected for \citet{S+S73} disks,
although discrepancies of order a factor of a few have been reported
\citep{Poo++07}. Assuming that the size scales as $\lambda^{1/\eta}$,
$\eta$ is found to be in the range $0.5-1$, whereas $\eta=0.75$ is
expected for a \citet{S+S73} disk
\citep{PMK08,Eig++08b}. Long wavelength data imply the presence of a
second spectral component, consistent with the hypothesis of a dusty
torus of size much larger than the microlensing scale \citep{Ago++09}.

These first exciting results are just the beginning, because very few
light curves obtained so far are long enough to harness the full power
of microlensing. With the rapid development of time-domain astronomy
predicted for the next decade, multiwavelength monitoring campaigns of
several years for tens of objects should become feasible
(\S~\ref{sec:future}).

\subsection{Cosmic telescopes and human telescopes}

I have described how strong lensing provides a unique opportunity to
study sources that are too faint or too small to be studies otherwise,
from quasar host galaxies to microarcsecond size accretion disks.

Unfortunately, the use of galaxies (and clusters) as cosmic telescopes
is often more contentious than it should be. One frequent critique is
that source reconstruction is difficult and inherently uncertain. This
is a false perception. The brief discussion in~\S~\ref{sec:theory} and
the references listed therein provide ample documentation that lens
modeling is now a mature field with very well understood
uncertainties, capable of delivering results that are well reproduced
by independent analyses. Lens modeling at cluster scales is more
complex due to the larger dynamic range in the data and the more
inhomogeneous mass distribution. Nevertheless, robust results can be
obtained also for clusters, provided that enough information is
available.

Another frequent critique is that surveys using cosmic telescopes are
inefficient compared to blank fields because of magnification
bias. This is true for sources with number counts in flux units
(dN/dF) flatter than F$^{-1}$. However, when probing the bright-end of
the luminosity function of any population -- where number density
falls off exponentially -- lensing is just unbeatable: the brighter of
any class of distant astronomical objects will inevitably be
gravitationally lensed. Cosmic telescopes and blank surveys are
complementary to fully characterize a source population and its
physical properties.

\section{Searches for Gravitational Lenses}
\label{sec:searches}

The strong lensing applications covered in this article span a broad
range of astrophysical phenomena, observational, and theoretical
challenges. However, they all share a common limitation: the
relatively small number of systems to which they can be
applied. Although there are 200 systems known, they are not all
suitable for all applications. Studies must rely on at most a few tens
of cases to infer results of general interest.

Fortunately, a number of large surveys are expected to take place in
the next decade, providing an ideal dataset to mine for rare objects
such as strong lenses. The challenge will consist in developing fast
and robust algorithms to find new lenses, and then in mustering the
resources and the brain power needed to follow them up and study them
(\S~\ref{sec:future}).

Before I summarize some of the searching techniques, it is useful to
establish a discovery ``etiquette'': what are the necessary and
sufficient elements to identify a strong gravitational lens? Here are
two necessary criteria: i) multiple images clearly identified; ii)
image configuration reproduced by a ``simple'' model. The first
criterion seems to me unavoidable, although it has not always been
applied in the past. The second criterion is more subjective, but can
be made quantitative in the following way. Given our knowledge of the
surface brightness distribution of galaxies and of the gravitational
potential, is it more likely that the observed configuration arises
from some random configuration (e.g., HII regions distributed along a
cross-pattern, or two quasars with similar colors on the opposite
sides of a galaxy), or from strong lensing of a more common surface
brightness distribution? It seems to me these two criteria are also
sufficient. Additional criteria such as images having identical colors
or spectroscopic redshift of deflector and source are desirable, but
impractical for future surveys that may have high resolution images in
just a single band, or limited capabilities for spectroscopic
follow-up.

\subsection{Imaging-based searches}

Imaging-based searches can be divided into catalog-based and
pixel-based. Catalog-based searches look for objects in a lensing-like
configuration. They are most effective at detecting sharp
multiply-imaged features such as multiply-imaged quasars
\citep[e.g.][]{Ina++08,Ogu++08}, but they can also be used for
extended sources, provided the image separation is large enough for
deblending \citep[][]{Bel++07a,All++07}. Pixel-based searches start
from a set of pixels, and look for lensing-like configurations. Lenses
are identified on the basis of characteristic geometries
\citep[e.g.][]{Cab++07} or by actually modeling every system as a
possible lens \citep{Mar++09}. The pixel-based method is slower and
more computationally intensive than catalog-based searches, but in
principle can be used to push the detection limit to smaller angular
separations, beyond the level where source and deflector can be
deblended by general-purpose cataloging softwares. Visual searches can
be considered as pixel-based, with the human brain as lens-modeling
tool \citep[e.g.][]{NMT09,Jac08}. Algorithms need to be tweaked to
reach an optimal balance between completeness (false negative) and
purity (false positive) appropriate for each dataset and scientific
goal. The best algorithms can currently achieve 90\% completeness and
purity searching through HST data \citep{Mar++09,NMT09}. Although some
human intervention is still necessary, this breakthrough makes it
feasible to search through future surveys of 1000 deg$^2$ or more.
 
Time-domain surveys allow for a different image-based strategy:
looking for variable resolved sources \citep{Koc++06b}. At high
galactic latitude, lensed quasars are more common than contaminants
such as pairs of variable stars. Pairs of non-lensed quasars can be
distinguished on the basis of their light curves and colors, while
lensed supernovae are a welcome contaminant (see
\S~\ref{sec:future}). A first application of the method to the SDSS
Supernovae survey data show that the only known compelling lens
candidate is recovered as a close pair of variable sources. Out of
over 20,000 sources, only a handful of false positives are found,
suggesting a ``purity'' of $\sim$20\% \citep{Lac++09}. This is
encouraging, although more tests on wider and deeper data are needed
to further improve the method in view of upcoming surveys.

\subsection{Spectroscopy-based searches}

Spectroscopic searches rely on identifying composite spectra with
features coming from multiple redshifts. Follow-up high resolution
information is then needed to identify the subset of events with
detectable multiple-images, and to obtain astrometry for lens
modeling. A strong advantage of the method is that lenses come with
redshifts by construction. After the early serendipitous discoveries
\citep{Huc++85}, the method started to bear large numbers of lenses
only with the SDSS spectroscopic database
\citep{Bol++06,Wil++06,Bol++08a,Aug++09a}.  The recent searches
highlight the quality of spectroscopic data as the key element for
success. High signal-to-noise ratios are needed to identify faint
spectral features, close-to Poisson limited sky subtraction is needed
to reduce false positives, spectral resolution better than 100 \kms\,
is need to resolve line multiplets, and wide wavelength coverage
increases the redshift range for the search. It is a testament to the
high quality of the SDSS database that the confirmation rate is $\sim
60-70$ \% \citep{Bol++08a}, after a very strict initial selection
(approximately 1/1000 SDSS galaxies are selected as a candidate for
follow-up by SLACS).

\section{Future Outlook}
\label{sec:future}

\subsection{Thousands of gravitational lenses}

Most of the applications listed in the previous sections are listed by
sample size. An increase by one of order of magnitude in sample size
is needed to make progress. Fortunately, there is a realistic
opportunity to make this happen in the next decade, considering the
typical yields for strong lens systems searches.  For optical and near
infrared imaging searches, yields are $\sim$10 deg$^{-2}$ at HST-like
depth and resolution \citep{MBS05}, $\sim$1 deg$^{-2}$ at the best
ground based conditions \citep{Cab++07}. At radio wavelengths and
$0\farcs25$ resolution expected for the Square Kilometer Array
\citep{Koo++09b} the yield is $\sim$1 deg$^{-2}$.  For spectroscopic
surveys, the yield is $\sim10^{-3}$/spectrum.  Thus, a 1000 deg$^2$
HST-quality cosmic shear survey, all sky ground based surveys in the
optical or radio, and a 10$^7$ galaxy redshift survey should all be
capable of yielding $\sim 10,000$ strong gravitational lens systems,
although with different properties. High angular resolution surveys
will be critical for applications such as the study of small mass
deflectors and of the substructure mass function. Time-domain surveys
will have a built-in advantage for, e.g., time-delays and
microlensing. Spectroscopic surveys will be advantageous for those
applications that require redshift and velocity dispersions, such as
the study of luminous and dark matter in the deflector. Several
thousand strong lens systems from each of these search techniques is
an ambitious, yet feasible, goal for the next decade.

These massive undertakings will require large number of people and
resources. As in many other instances, it is likely that such projects
will require the joint efforts of a number of communities interested
in diverse scientific questions. The unique capabilities of strong
lensing make it very worthwile to design future surveys keeping in
mind its requirements. 

\subsection{The problem of follow-up}

Let us assume that 10,000 strong lens candidates have been found. What
follow-up will be needed to extract scientific information? Images
with resolution of order $0\farcs1$ are often key to prove the lensing
hypothesis, and to construct detailed lens models and study the
properties of the host and the source. If the resolution of the finder
survey is not adequate, follow-up will be required. Current follow-up
imaging typically requires an orbit of HST. JWST should gain in speed
for most applications and be revolutionary for long-wavelength
studies, such as flux ratio anomalies. For a subset of objects with
suitable colors and nearby stars, high resolution imaging could
perhaps also be obtained in a comparable amount of time with a 8-10m
telescope equipped with laser guide star adaptive optics
(LGSAO). Extremely large 30-m class telescopes (ELTs) with LGSAO
should be able to gain a substantial factor in speed and
resolution. Radio follow-up of extended sources at high resolution
with VLA requires of order 1 hour per lens. The Atacama Large
Millimiter/Submillimiter Array (ALMA) should be an improvement both in
speed and resolution. Even following up a thousand lenses will thus
require thousands of hours of telescope time, maybe a few hundreds
with JWST, ELTs and ALMA. This may be feasible, but not trivial,
making high resolution imaging a likely bottle neck. Multiplexing is
unlikely to be an option given the rarity of these objects on the sky,
although multiplexing with different astronomical targets is certainly
a desirable option.  Even higher resolution images ($0\farcs01$) are
within reach with extreme adaptive optics on extremely large
telescopes and will certainly be beneficial for pushing some of the
lensing applications. For example, that kind of resolution could push
the detection of DM substructure in distant galaxies in the 10$^7$
M$_{\odot}$ regime typical of the least massive luminous Milky Way
satellites currently known, where the discrepancy with theory is
currently strongest \citep{Kra10}.

Spectroscopic follow-up to gather redshifts is a problem of possibly
even greater magnitude, considering that redshifts for many of the
sources cannot be measured even spending hours on the largest
telescopes \citep{Ofe++06}.  For the fainter sources, photometric
redshifts may be the only option. Coordination with redshift surveys
-- such as those proposed to measure baryonic acoustic oscillations --
will help measuring redshifts as well as in spectroscopic searches,
although they will also require high angular resolution follow-up.

Monitoring campaigns of thousands of lensed AGNs are out of the
question at the moment, but could be a natural byproduct of future
synoptic surveys. Some of the most demanding time domain applications,
such as detection of time-delay anomalies could be beyond the reach of
ground based monitoring tools and require a dedicated space mission
\citep{Mou++08}.

In parallel with discovery efforts, careful thought must be put into
planning follow-up efforts. First, ways to extract as much information
as possible from the discovery images themselves must be
found. Second, follow-up efforts should be coordinated as much as
possible with those of other science cases to find common paths and
synergies. Last but not least, brain power could be another serious
limitation. Currently, accurate and reliable lens models require
several days of expert human brain activity. This will not be possible
when samples will consist of tens of thousands of systems.

\subsection{Unusual lensing applications in an era of abundance}

I conclude with four examples of strong lensing applications that
require very rare conditions and therefore need the large samples
expected in the next decade to become viable.

Lensed supernovae Ia are extremely valuable, because their standard
luminosity constrains the absolute magnification and therefore breaks
the mass-sheet degeneracy. For typical rates, we expect of order one
could be found monitoring known lenses. However, a ground based
time-domain survey covering most of the sky is expected to find of
order a hundred lensed type Ias (Oguri \& Marshall, 2009, in
preparation).

Compound lenses are potentially powerful cosmographic probes, but
there is currently only one such system known at galaxy scales
\citep{Gav++08}. Thousand square degree field surveys at HST-like
resolution should be able to find tens of systems like SDSSJ0946+1006,
potentially constraining $w$ to the 10 \% level \citep{Gav++08}.

Strong lensing is one of the few tools capable of measuring the mass
of quiescent black holes at cosmological distances, through their
gravity affects the properties of central images
\citep{MWK01}. Detecting the central image -- which is generally
highly demagnified -- is usually beyond reach with current
instrumentation \citep[see however][]{WRK04}. However this application
may become practical with future facilities, especially at radio
wavelengths where the contrast between deflector and source is more
favorable.

Finally, with future samples of $10^4$ lenses, rare examples of
``catastrophes'' should be identifiable \citep{O+M09}. These are very
special lensing configurations characterized by specific constraints
on the gravitational potential and its derivatives, and they only
occur only for very specific source position and redshift
\citep[see][for details]{SEF92,PLW01}.  The identification of
examples of catastrophes is interesting for two reasons. Firstly,
catastrophes often lead to extreme magnification factors, up to $\sim
100$, making them extraordinary cosmic telescopes. Secondly, the
unusual geometry of multiple images can give remarkably strong
constraints on the mass distribution of the deflector \citep{O+M09}.

\section*{Acknowledgments}

I am grateful to Matt~Auger, Maru\v{s}a~Brada\v{c}, Brendon~Brewer,
Chris~Fassnacht, Raphael~Gavazzi, Leon~Koopmans, Phil~Marshall,
Elisabeth~Newton, Carlo~Nipoti, Andrea Ruff, Anna~Pancoast, and Sherry
Suyu for helpful comments on early drafts of this article. I would
like to thank in particular Richard~Ellis for his many insightful
comments and advice about this article and throughout the past
decade. I am also grateful to Roger~Blandford, for his scientific
input and feedback as editor. Finally, I would like to acknowledge
enlightening conversations with many people including Giuseppe~Bertin,
Chris~Kochaneck, Jean-Paul~Kneib, Paul~Schechter, Peter~Schneider, and
Joachim~Wambsganss. Financial support from the Sloan and Packard
Foundations is gratefully acknowledged.

\section*{Acronyms}

\begin{enumerate}
\item CDM: Cold Dark Matter
\item CMB: Cosmic Microwave Background
\item DM: Dark Matter
\item HST: Hubble Space Telescope
\item IMF: Initial Mass Function
\item JWST: James Webb Space Telescope
\item SDSS: Sloan Digital Sky Survey
\item SIE: Singular Isothermal Ellipsoid
\item SIS: Singular Isothermal Sphere
\item SLACS: Sloan Lens Advanced Camera (for Surveys) Survey
\end{enumerate}

\section*{Definitions}%

\begin{enumerate}

\item Convergence: dimensionless projected surface mass density in units of the critical density.

\item Deflector: the foreground galaxy responsible for the lensing potential.

\item Einstein radius: characteristic scale of strong lensing.
For a circular deflector it corresponds to the radius within which
$\langle \kappa \rangle = 1$.

\item Image plane: twodimensional map of the source emission as it appears to the observer after propagation through the lensing potential.

\item Macrolensing: strong lensing producing image separations of order arcseconds, the typical scale of massive galaxies.

\item Microlensing: strong lensing producing image separation of order of micro-arcseconds, the typical scale of individual stars.

\item Millilensing: strong lensing producing image separation of order of milli-arcseconds, the typical scale of small satellite galaxies.

\item Shear: dimensionless quantity that describes the local distortion of lensed images.

\item Source: the background astronomical object whose light is being lensed.

\item Source plane: twodimensional map of the source emission as it would appear to the observer in the absence of a deflector.

\item Strong lensing: deflection of light from a background source by a foreground deflector strong enough to produce multiple images.

\end{enumerate}

\section*{Summary Points}

\begin{enumerate}

\item Massive early-type galaxies are surrounded by dark matter halos spatially more extended than the luminous component. The fraction of mass in the form of  dark matter inside the effective radius increases with galaxy stellar mass.

\item The total mass density profile of massive early-type galaxies is approximately isothermal in the innermost $\sim$10 kpc, i.e. the logarithmic slope $\gamma'$ equals two within 10\%.

\item Precise gravitational time delays for a single system can be used to measure the Hubble Constant to 5\% precision, provided that enough information is available to constrain the local gravitational potential and to break the mass-sheet degeneracy. Time delays break the degeneracy between $h$ and $w$ in the analysis of CMB data. Combining the constraints from the lens system B1608+656 and those from WMAP5 yields $h=0.697^{+0.049}_{-0.050}$ and $w=-0.94^{+0.17}_{-0.19}$ assuming flatness.

\item The host galaxies of distant luminous quasars appear to be underluminous in comparison with local galaxies hosting black holes of the same mass. This may indicate that in this mass range black holes complete their growth before their host galaxy.

\item Microlensing results indicate that the size of accretion disks and its dependency on temperature is in broad agreement with the predictions of \citet{S+S73} models. Moreover, mid-infrared microlensing studies are consistent with a presence of an unresolved dusty region, larger than the accretion disk.

\end{enumerate}

\section*{Future Issues}

\begin{enumerate}
\item How do luminous and dark matter density profiles depend on galaxy mass, type, and cosmic time? 

\item Are dark matter density profiles universal, as predicted by CDM numerical simulations? 

\item Is the mass function of substructure in agreement with the predictions of CDM numerical simulations?

\item How are density profiles and the substructure mass function influenced by the presence of baryons?

\item Is dark energy the cosmological constant ($w=-1$)? If not, how does the equation of state evolve with cosmic time?

\item How can we find and exploit larger samples of strong gravitational lens systems?

\end{enumerate}

\begin{figure}
\centerline{\psfig{file=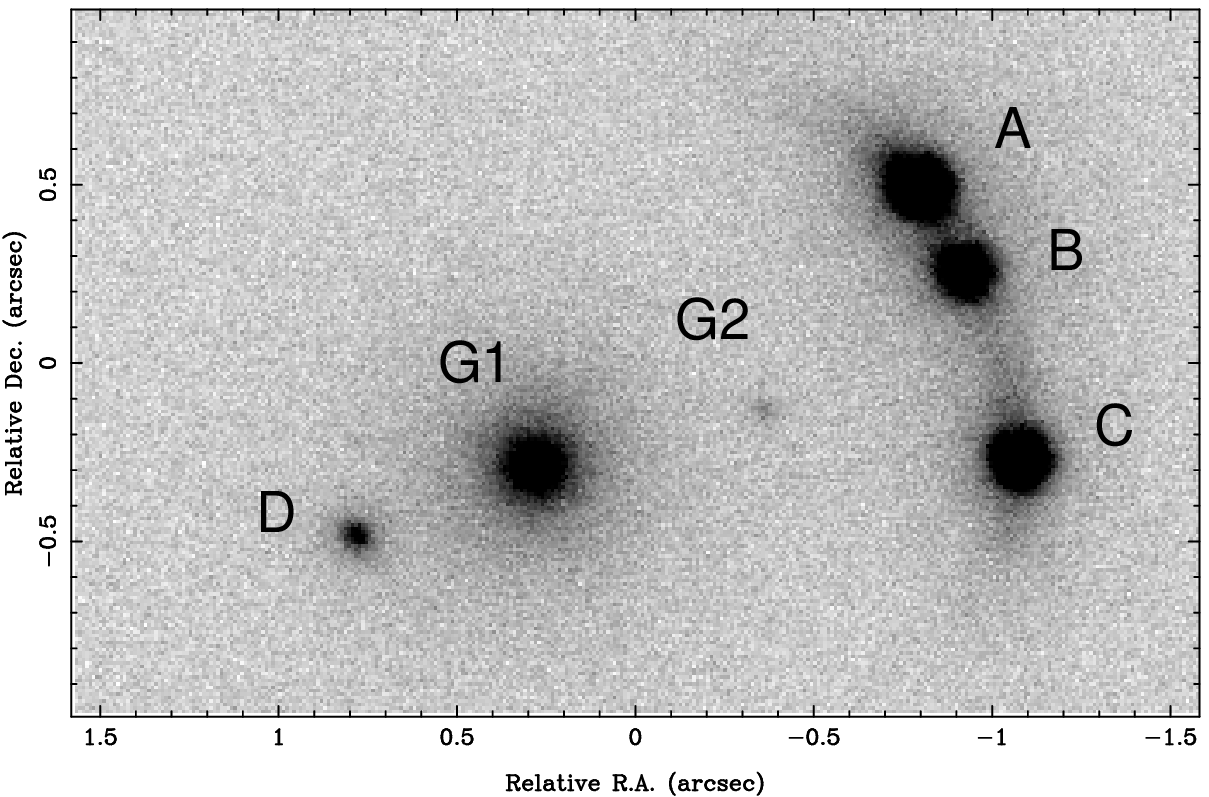,width=\textwidth}}
\caption{Near infrared (2.2$\mu$m) image of the gravitational lens
system B2045+265 taken with the adaptive optics system at the 10m
Keck-II Telescope \citep[from][]{McK++07}. For this kind of
configuration, the flux of image B is expected to be equal to the sum
of the fluxes of images A and C, in the absence of substructure
\citep[][]{Bra++02,KGP03}.  The anomaly was originally discovered on
the basis of radio images, ruling out microlensing or differential
interstellar medium scattering as alternative interpretation
\citep[][]{Fas++99b,Koo++03a}. 
A satellite galaxy (G2) of the main deflector (G1) is detected in this
deep and high resolution image. A small mass located at the position
of the satellite can explain the observed anomaly.}
\label{fig:anomaly}
\end{figure}

\section{LITERATURE CITED}



\end{document}